\PassOptionsToPackage{protrusion, expansion}{microtype} %% for better typography
\PassOptionsToPackage{dvipsnames}{xcolor} %% for better typography
\pdfoutput=1
\documentclass[conference, nofonttune]{IEEEtran}\def\UseIEEETemplate{1}  %% my macro
\IEEEoverridecommandlockouts%% intended to improve IEEE author block

\usepackage{graphicx}
\usepackage{xcolor}
\definecolor{B}    {HTML}{2b66d3}   %% blue, light
\definecolor{B2}   {HTML}{003399}   %% blue, dark 
\definecolor{Bv}   {HTML}{0000EB}   %% blue, vibrant
\definecolor{R}    {HTML}{c9171e}
\definecolor{R2}   {HTML}{d7003a}
\definecolor{INK}  {HTML}{595857}
\definecolor{Y}    {HTML}{f1c40f}
\definecolor{G}    {HTML}{009a00}
\definecolor{GRAY} {HTML}{808080}
\definecolor{MAUVE}{HTML}{9400D1}

\usepackage{array}      %% https://tex.stackexchange.com/q/338009
\usepackage{booktabs}
\usepackage{colortbl}
\usepackage{adjustbox}
\usepackage{multirow}
\usepackage{tabu}       %% friendly when coloring

\usepackage{url, hyperref}                  %% hyperlink
\hypersetup{
    colorlinks, linkcolor=RoyalBlue, citecolor=RoyalBlue, urlcolor=RoyalBlue,
    allcolors=RoyalBlue,
    pdfborder={0 0 0}}

\usepackage{enumitem}                       %% enhanced items
\usepackage{lipsum}                         %% pseudo texts
\usepackage{xpatch}% specifically for fixing indentation, thanks to https://tex.stackexchange.com/questions/294840/indent-algorithmic-outside-of-a-block

\usepackage{listings}						%% code listing
\usepackage{setspace}                       % spacing

\usepackage{soul}%% \hl{} and \st{}
\ifx\UseACMTemplate\undefined
\else
    \newcommand{\SEC}{\textcolor{black}{\S}}

\fi

\ifx\UseIEEETemplate\undefined
\else
    \newcommand{\SEC}{\textcolor{black}{\S}}

\fi

\colorlet{HLCOLOR}{B}
\colorlet{TableAltColor}{gray!20}

\usepackage{tikz}
\usetikzlibrary{positioning}        %% must-have
\usetikzlibrary{shapes.multipart}   %% multiline node, should use as default
\usetikzlibrary{shapes.geometric}   %% diamond etc
\usetikzlibrary{arrows}
\usetikzlibrary{arrows.meta}        %% e.g., {[left]Bar}-{[right]Bar}
\usetikzlibrary{backgrounds}
\usetikzlibrary{patterns}

\usepackage{pgfplots}
\usepackage{pgfplotstable}
\usepgfplotslibrary{groupplots}

\usepackage{anyfontsize}

\usepackage{amsmath}    %% \text in mathmode
\usepackage{amsfonts}
\usepackage{amsthm}
\usepackage{amssymb}    %% may cause conflict
\usepackage{mathtools}
\usepackage{bm}

\usepackage{caption}
\usepackage{subcaption}
\usepackage[normalem]{ulem}
\usepackage{stackengine}
\usepackage{rotating}	%% dependency of turn

\usepackage[outercaption]{sidecap}
\sidecaptionvpos{figure}{b}
\sidecaptionvpos{table}{b}
\hypersetup{
	pdftitle={FT K-means: A High-Performance K-means on GPU
with Fault Tolerance},
	pdfauthor={Wu, Ding, et al.},
	pdfsubject={FT K-means, camera-ready},
	pdfkeywords={GPU; Fault Tolerance; Machine Learning; HPC}
}

\usepackage{amsmath}
\DeclareMathOperator*{\argmax}{arg\,max}

\usepackage{listings}
\usepackage{microtype}

\usepackage{enumitem}
\usepackage[switch]{lineno}
\usepackage{algorithm}
\usepackage{algpseudocode}
\usepackage{multicol}
\usepackage{multirow}
\usepackage{caption}
\usepackage{subcaption}
\usepackage{subcaption}
\usepackage{booktabs} % For nice tables
\usepackage{siunitx} % To align table numbers by unit
\usepackage{etoolbox}
\usepackage{soul}
\usepackage{cancel}
\usepackage[normalem]{ulem}

\usepackage{caption}

\usepackage{amsmath,algorithm,tabularx}
\usepackage{algpseudocode}

\makeatletter
\newcommand{\multiline}[1]{%
  \begin{tabularx}{\dimexpr\linewidth-\ALG@thistlm}[t]{@{}X@{}}
    #1
  \end{tabularx}
}
\renewcommand{\fnum@figure}{Figure \thefigure}
\newcommand{\shixun}[1]{\textcolor{black}{#1}}

\makeatother
\microtypesetup{nopatch={footnote}}

\newcommand{\EDIT}[2]{%
    \bgroup%
    \colorbox{B}{\color{white}#1:}\color{B} #2
    \egroup%
}

\usepackage[style=ieee, citestyle=numeric, sortcites=true, backend=bibtex]{biblatex}
\usepackage{algorithm}
\usepackage{algpseudocode}
\begin{document}
\title{DGRO: \underline{D}iameter-\underline{G}uided \underline{R}ing \underline{O}ptimization for Integrated Research Infrastructure Membership}

%% change when camera-ready
% \author{(author list placeholder: double-blinded)}
\newcommand{\CoMark}{$^{\star}$}
\newcommand{\UcrMark}{$^\dagger$}
\newcommand{\UHMark}{$^\S$}
\newcommand{\AnlMark}{$^\ddagger$}

\newcommand{\FsuMark}{$^\ddagger$}
\newcommand{\UiowaMark}{$^\S$}

\author{%
\normalsize\null
    Shixun Wu\UcrMark,
    % Yitong Ding\UcrMark\CoMark,\thanks{\CoMark\,Shixun Wu and Yitong Ding contributed equally to this work.}
    % Yujia Zhai\UcrMark,
    % Jinyang Liu\UHMark,
    % Jiajun Huang\UcrMark,
    % Zizhe Jian\UcrMark,
    % Huangliang Dai\UcrMark,\\
    Krishnan Raghavan\AnlMark,
    Sheng Di\AnlMark,
    % Bryan M. Wong\UcrMark,
    Zizhong Chen\UcrMark,
    Franck Cappello\AnlMark
    \\
    \IEEEauthorblockA{\UcrMark University of California, Riverside, CA, US}
    % \IEEEauthorblockA{\IuMark Indiana University, Bloomington, IN, USA}
    % \IEEEauthorblockA{\UHMark University of Houston, Houston, TX, US}
    \IEEEauthorblockA{\AnlMark Argonne National Laboratory, Lemont, IL, US}
    % \IEEEauthorblockA{\CoMark Co-first Authors}
    % \IEEEauthorblockA{\UiowaMark The University of Iowa, Iowa City, IA, USA}
    % \IEEEauthorblockA{\FsuMark Florida State University, Tallahassee, FL, USA}
    % swu264@ucr.edu, dingyitongytd@gmail.com, yzhai015@ucr.edu, jliu447@ucr.edu, jhuan380@ucr.edu, zjian106@ucr.edu, 
    swu264@ucr.edu, kraghavan@anl.gov, sdi1@anl.gov, chen@cs.ucr.edu, cappello@mcs.anl.gov
}

%% comment before submission; only for counting pages
\thispagestyle{plain}\pagestyle{plain}

\maketitle

% A single error in FFT leads to an exponential increase in the total number of errors. Due to the error propagation, existing fault tolerance schemes not only necessitate a checksum computation for each signal and a time-redundant recompute for error correction but also lack architecture-aware optimization on GPU. 
%%
%% Keywords. The author(s) should pick words that accurately describe
%% the work being presented. Separate the keywords with commas.
% \keywords{FFT, GPU, Performance Optimization, Reliability, Resilience}

%% A "teaser" image appears between the author and affiliation
%% information and the body of the document, and typically spans the
%% page.

%%
%% This command processes the author and affiliation and title
%% information and builds the first part of the formatted document.

% \settopmatter{printfolios=true}

% \ccsdesc[500]{Computing methodologies~Massively parallel algorithms}

%\IEEEtitleabstractindextext{%

\begin{abstract}	
Logical ring is a core component in membership protocol. However, the logic ring fails to consider the underlying physical latency, resulting in a high diameter. To address this issue, we introduce \textbf{D}iameter-\textbf{G}uided \textbf{R}ing \textbf{O}ptimization (DGRO), which focuses on constructing rings with the smallest possible diameter, selecting the most effective ring configurations, and implementing these configurations in parallel. We first explore an integration of deep Q-learning and graph embedding to optimize the ring topology. We next propose a ring selection strategy that assesses the current topology's average latency against a global benchmark, facilitating integration into modern peer-to-peer protocols and substantially reducing network diameter. To further enhance scalability, we propose a parallel strategy that distributes the topology construction process into separate partitions simultaneously. Our experiment shows that: 1) DGRO efficiently constructs a network topology that achieves up to a 60\% reduction in diameter compared to the best results from an extensive search over \(10^5\) topologies, all within a significantly shorter computation time, 2) the ring selection of DGRO reduces the diameter of state-of-the-art methods Chord, RAPID, and Perigee by  10\%-40\%, 44\%, and 60\%. 3) the parallel construction can scale up to $32$ partitions while maintaining the same diameter compared to the centralized version. 
\end{abstract}

% Contribution:
% 1. Job membership idea 
% 2. topology diameter optimization
% 3. parallel q-learning method
% 4. scalability, partial DHT
% 5. fault tolerance 
% 5. Efficient program and data migration

%\begin{IEEEkeywords}
%	scientific data base, error-bounded lossy compression, data reduction, large-scale scientific simulation
%\end{IEEEkeywords}
%}
\section{Introduction}

Researchers increasingly require cross-platform utilization of diverse computing and storage devices \cite{none2021toward,liu2024high,jian2024cliz,liu2023cusz,huang2023exploring}. The U.S. Department of Energy (DOE) recognizes the persistent trend of integrated research infrastructure (IRI) as the future of scientific computing \cite{brown2021vision}. These intricate science workflow processes utilize geographically dispersed computational resources across multiple facilities \cite{ahrens2022envisioning}. However, the diversity and distributed nature of these resources, managed by different organizations, domains, and communities, present significant challenges in fully leveraging their capabilities \cite{bard2022lbnl}. Consequently, effective membership management and reliable failure detection are crucial for the stability and efficiency of these systems.

Moreover, at such scales, system failures are commonplace rather than exceptions \cite{barroso2022datacenter,dean2013tail}. Researchers encounter numerous issues, including application code errors \cite{stelling1999fault}, authentication issues, network disruptions \cite{losada2020fault}, workflow system breakdowns \cite{stelling1999fault}, filesystem and storage complications \cite{patterson1988case}, and hardware faults \cite{sridharan2012study,park2009reliability,wu2023anatomy,wu2023ft,wu2024turbofft,wu2024ft}. Enhancing the computational platform’s performance and resilience making it fault-tolerant, robust against a variety of workloads and environmental conditions, and adaptable to changes in workloads, resource availability, and connectivity is crucial for enabling researchers to pursue their scientific goals efficiently and effectively \cite{cappello2014toward}.

% Effective membership protocol management within an Integrated Research Infrastructure (IRI) is crucial for optimizing resource allocation and enhancing the performance and reliability of scientific workflows. In environments where resources are distributed across different organizations and domains, each with their own management systems and operational policies, the role of a well-designed membership protocol becomes even more significant. Such protocols facilitate coherent resource management by standardizing interactions and data exchanges among diverse systems, thereby ensuring that resources are utilized efficiently and scientific workflows are executed smoothly. This standardization helps in mitigating conflicts and reducing overhead associated with the coordination of disparate systems, ultimately leading to improved throughput and reduced latency in scientific computations. In essence, effective membership protocol management not only streamlines resource allocation but also serves as a backbone for maintaining the integrity and efficiency of the entire research infrastructure.

Membership management solutions in modern systems can generally be categorized into two types, centralized or decentralized gossip-based. The first is a centralized strategy, such as Slurm \cite{yoo2003slurm}, etcd \cite{etcd2014}, or Chubby \cite{burrows2006chubby}, or ZooKeeper \cite{zookeeper2010apache}, to manage the membership list. This approach offers simplicity by maintaining a centralized list with strong consistency, which other processes periodically access. However, this centralization can limit system resilience, as failures or connectivity issues within this small cluster could compromise service availability \cite{netflix2014,kelley2014eureka,zookeeper2010apache}. The second category comprises gossip-based, fully decentralized methods \cite{hewitt2010cassandra,akka2009,redis2009,newell2016optimizing,suresh2018stable,scylladb2013}, which offer enhanced resilience by distributing the membership information across the network, thereby reducing reliance on a central point of failure. Gossip-based membership. van Renesse et al. \cite{van1998gossip,van2003astrolabe} was proposed to handle membership by utilizing gossip to disseminate positive notifications (keepalives) among all processes. If a process \( p \) crashes, other processes will eventually remove \( p \) after a timeout period. SWIM \cite{das2002swim} introduced a variation of this method, which minimizes communication overhead by using gossip to distribute “negative” alerts instead of the usual positive notifications. Nowadays, gossip-based membership protocols are commonly implemented in various deployed systems, such as ScyllaDB \cite{scylladb2013}, Akka \cite{akka2009},  Redis Cluster \cite{redis2009},Cassandra \cite{hewitt2010cassandra},  Orleans \cite{newell2016optimizing}, Netflix’s Dynomite \cite{netflix2014}, Uber’s Ringpop \cite{suresh2018stable}, and systems at Twitter \cite{mccaffrey2015}.

\begin{figure}[h!]
    \vspace{-3mm}
    \centering
    \includegraphics[width=\linewidth]{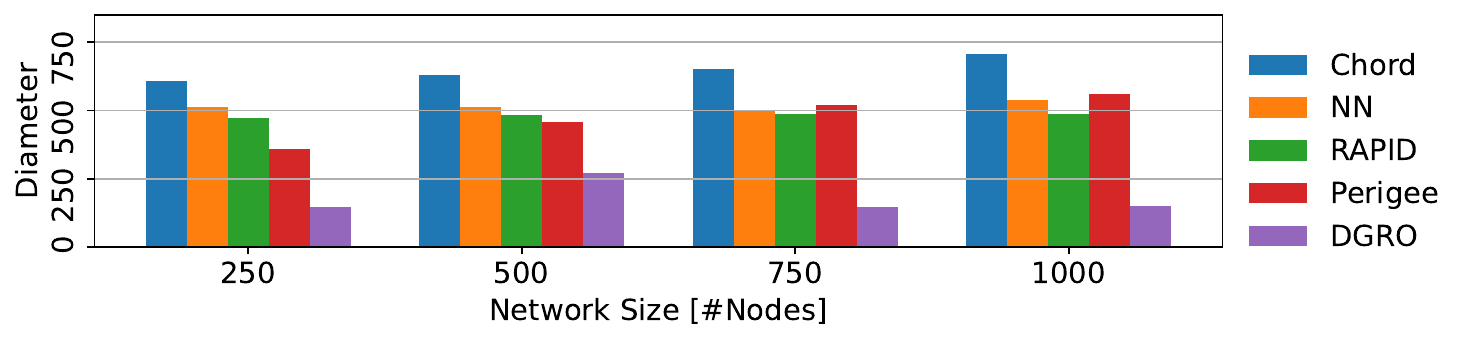}
    \caption{DGRO has low diameter.}
    \label{fig:DGRO_low_diameter}
    \vspace{-5mm}
\end{figure}
Decentralized gossip protocols commonly utilize a logical ring, structured by consistent hashing, which is inherently random and disregards the physical network layout. This randomness often leads to inefficiencies, as the ring fails to minimize network diameter and thus exacerbates latency issues. As shown in Figure \ref{fig:DGRO_low_diameter}, state-of-the-art has a diameter up to 3 times higher than our DGRO.

% Today's resource management infrastructure often relies on centralized architectures, where workflow management systems (WMS) and cluster schedulers need to have a global view of the target resources~\cite{deelman2021pegasus,mitchell2019exploration,harenslak2021data,tyler2022cross,svirin2019bigpanda,galaxy2022galaxy,alsaadi2022radical,bockelman2021principles,frey2002condor,wratten2021reproducible,yoo2003slurm}. Such centralized architectures suffer from a lack of resilience, efficiency, and scalability. The SWARM project takes a novel viewpoint, where distributed intelligence, specifically swarm intelligence (SI)~\cite{Kennedy2006,bonabeau1999swarm} is coupled with a rigorous multi-agent optimization framework to enable large-scale, efficient and resilient resource management for scientific workloads. 

% The induced peer-to-peer (P2P) network topology can often be highly suboptimal because it typically overlooks the geographical distances between peers. This oversight can lead to inefficient network configurations where data must travel longer paths than necessary, increasing latency and reducing the overall speed of data transmission across the network. In scenarios where real-time data processing and quick response times are critical, such inefficiencies can significantly hinder the performance and reliability of distributed systems. By not factoring in the physical proximity of nodes, the network may not only suffer from slower data exchanges but also from increased costs and energy consumption associated with data transmission over extended distances.

In contrast, we address the inefficiencies of suboptimal ring topologies through our Diameter-Guided Ring Optimization (DGRO) approach, which tackles the problem from three key perspectives: 1) determining the minimum possible diameter for a given fixed degree, 2) selecting the most effective ring topology based on a given latency distribution, and 3) exploring the feasibility of constructing ring topologies in parallel without compromising the overall diameter, thus aiming to reduce the time consumption associated with sequential builds.
% determine introduces a revolutionary diameter-guided membership topology for integrated research infrastructure, distinguished by its use of K Hamilton Cycles, which consistently achieve a lower network diameter compared to traditional peer-to-peer methods that often overlook geographical latency impacts. Unlike existing solutions that mainly aim to reduce hop counts, our approach addresses the critical latency issues by integrating Q-learning with Graph Neural Networks (GNN) to optimize the construction of K-Hamilton Cycles. This method improves node selection by estimating future states rather than just relying on immediate gains, and our use of GNN allows the model to scale to variable membership sizes. Additionally, we've developed a parallel strategy that significantly accelerates the topology construction process, enhancing the speed and efficiency of the learning phase. This innovative approach not only minimizes network diameters but also addresses the inefficiencies of traditional P2P topologies that do not consider physical distances between nodes.

We summarize our contributions as follows:

\begin{itemize}[leftmargin=1.3em]
    \item We develop, DGRO, a diameter-guided ring optimization for integrated research infrastructure membership. DGRO is characterized by a ring optimization with a hybrid strategy of human heuristics and reinforcement learning, achieves a low diameter compared to other state-of-the-art methods.
    \item We design a Q-learning method with graph embedding to build a ring topology with low diameter. Q-learning is used to avoid local optima and greedy selection commonly found by human heuristics. This method selects nodes by utilizing a q-function to estimate future states rather than relying on a one-step greedy algorithm, minimizing the network diameter.
    \item DGRO dynamically selects an appropriate ring topology by comparing the average latency of the current topology against a range established by the global minimum and average latencies. This self-adaptive strategy can be integrated into contemporary peer-to-peer protocols, leading to a significant reduction in network diameter.
    \item We propose a parallel strategy to enhance the topology construction process, achieving the same diameter as the sequential build while speeding up by 32x.
    \item Experimental results show that: 1) DGRO efficiently constructs a network topology that achieves up to a 60\% reduction in diameter compared to the best results from an extensive search over \(10^5\) topologies, all within a significantly shorter computation time, 2) the ring selection of DGRO reduces the diameter of state-of-the-art methods Chord, RAPID, and Perigee by  10\%-40\%, 44\%, and 60\%. 3) the parallel construction can scale up to $32$ partitions while maintaining the same diameter compared to the centralized version.
    
\end{itemize}

The rest of this paper is arranged as follows: \SEC\ref{sec:background} introduces background and related works. \SEC\ref{section:system_model} describe the system model and performance metrics for our research. \SEC\ref{sec:DGRO} demonstrates the framework of DGRO. The design of self-adaptive ring selection is detailed in \SEC\ref{section:ring_topology_matters}, and the design of parallel ring construction is proposed in \SEC\ref{sec:parallel}. In \SEC\ref{sec:results}, the evaluation results are presented and analyzed. Finally, \SEC\ref{sec:conclusion} concludes and discusses future work.
  
\section{Background and Related Work}\label{sec:background}

\subsection{Physical Diameter Matters}

The diameter of a graph is the longest shortest path between any two vertices, representing a critical metric in network design, such as in information, social, and communication networks. It measures the maximum distance for direct communication within the network. A smaller diameter is vital for enhancing network efficiency by reducing latency in systems like multicore processor networks \cite{benini2002networks} or by creating highly influential networks for campaigns in small-world network models \cite{laoutaris2008bounded}.

Extensive research has focused on reducing the diameter of undirected graphs by adding new edges, a method commonly referred to as "shortcutting" in scholarly literature \cite{demaine2010minimizing,meyerson2009minimizing,tan2017shortcutting,chepoi2002augmenting}. These added edges are known as shortcut edges. A specific challenge in this area, the Bounded Cardinality Minimum Diameter (BCMD) problem, was formulated by Li, McCormick, and Simchi-Levi \cite{li1992minimum}. The objective in BCMD is to add a limited number of shortcut edges, typically no more than $k$, to reduce the overall diameter of the graph efficiently.

A notable limitation of the Bounded Cardinality Minimum Diameter (BCMD) problem is that optimal solutions, or their approximations, may significantly increase the degree of a single vertex. This increase is a typical outcome from prevalent approximation algorithms for BCMD, which often involve dividing the graph into \(k + 1\) distinct clusters and then connecting these clusters through a method called star-shortcutting. This process involves selecting a central vertex from one cluster and connecting it to the centers of all other clusters, consequently adding up to \(k\) shortcut edges and increasing the degree of the central vertex by \(k\).
\begin{figure}
    \centering
    \includegraphics[width=0.9\linewidth]{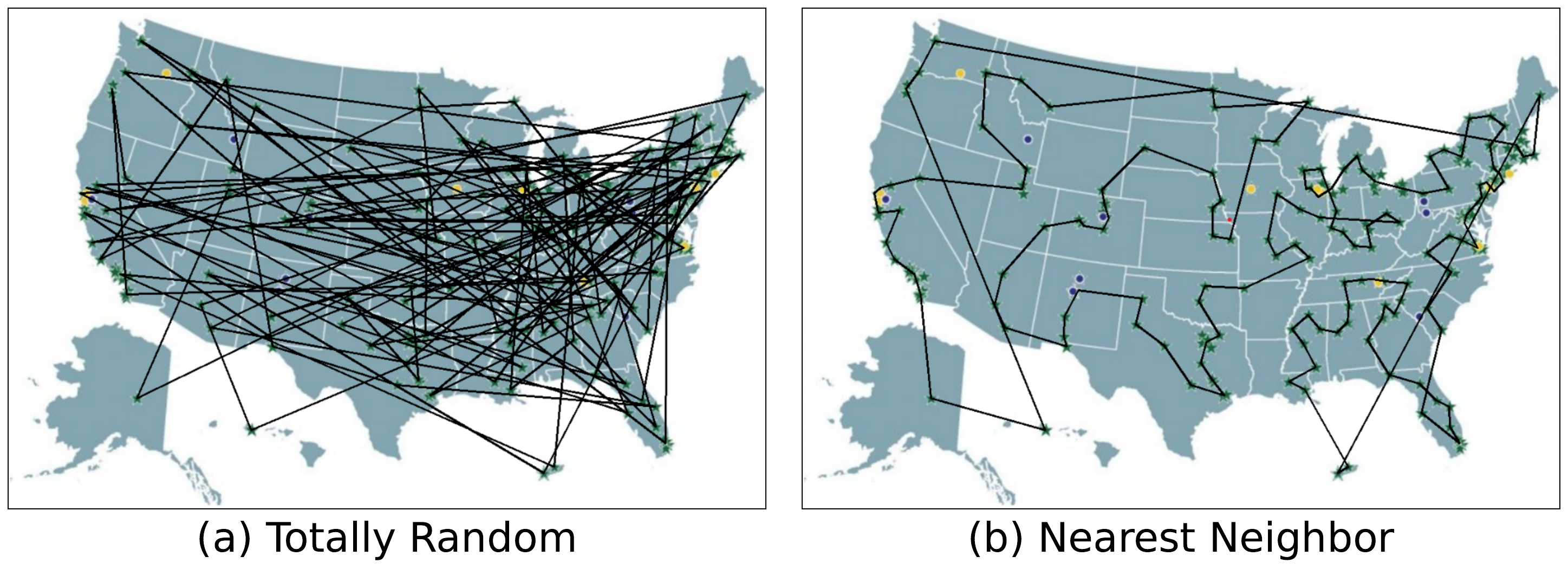}
    \caption{117 research sites located within U.S. Both (a) random ring and (b) nearest neighbor ring may introduce inefficient long jump between two geologically close nodes.}
    \label{fig:random_vs_nn}
    \vspace{-5mm}
\end{figure}
However, for many practical applications, it is important to limit the degree increase of any single vertex due to physical, economic, or other practical constraints. These constraints are commonly found in various real-world network challenges \cite{tan2017shortcutting,chung1984diameter}. For instance, Bokhari and Raza \cite{bokhari1986reducing} explored methods to decrease the diameter of computer networks by adding additional connections while ensuring that no more than one I/O port is added to each processor. Another scenario is in social networks, where service providers may want to suggest new friendships to enhance network connectivity and facilitate information spread while reducing polarization \cite{garimella2017reducing,haddadan2021repbublik,interian2021polarization}. In such cases, as noted in \cite{interian2021polarization}, the largest distance between members of different groups—termed the colored diameter—serves as a polarization metric. A more prudent approach might involve restricting the number of new connections recommended to each user to prevent overwhelming them and to ensure the practical materialization of these recommended links. 

In contrast to the above methods, we focus on the construction of one or multiple ring topologies that achieve a low diameter under degree constraints. In Figure \ref{fig:random_vs_nn}, both the random and nearest neighbor include long paths even between adjacent nodes.

\subsection{Graph Embedding\& Deep Q-learning}

The recent advancements in using deep learning architectures for solving combinatorial problems, such as the Traveling Salesman Problem (TSP) \cite{vinyals2015pointer,bello2016neural,graves2016hybrid,khalil2017learning,johnston2023curriculum,liu2023stationary,johnston2023downlink}, highlight the potential of the combination of reinforcement learning and graph embedding. These techniques are particularly effective in addressing the inherent complexities of combinatorial optimization. While existing architectures have struggled with efficiently capturing the combinatorial structure of graph problems and required extensive training data \cite{bello2016neural}, reinforcement learning and graph embedding can provide more tailored and efficient solutions. 

Moreover, the TSP shares similarities with ring optimization tasks, where the objective is to construct optimal loops, akin to finding the shortest possible route that visits each city once in TSP. Although both involve the construction of cycles, the primary difference lies in the objective or reward function used to evaluate solutions. In TSP, the goal is to minimize the total traveling distance, whereas ring optimization is to minimize the overall diameter. By adapting the methodologies developed for TSP, using reinforcement learning and graph embeddings, one can potentially find a better topology in terms of diameter.

\section{System Model}
\label{section:system_model}
\subsection{Network Model}
We model the integrated research infrastructure (IRI) peer-to-peer (P2P) network as an undirected graph \( G(V, E) \), where \( V \) represents the set of nodes and \( E \) denotes the set of edges, or links, between these nodes. A node corresponds to master machines (e.g., SLURM controller daemons) that can accept inter-cluster TCP connection requests from other servers and compute nodes. In contrast, compute nodes in the IRI are dedicated computing devices that cannot accept external TCP connection requests. Once a TCP connection is established between two nodes, communication flows bidirectionally.

In this work, we focus on the controller nodes, which form the core of the P2P overlay network. These nodes are typically always active, and the latency of membership changes is highly influenced by the interconnection network between them. Our objective is to minimize inter-cluster synchronization latency, not data transmission latency, a goal that is formally defined in §2.2. Our proposed topology optimization approach is generalizable and can be adapted to improve synchronization times for compute nodes as well.

For any pair of nodes \( (u, v) \) in the set \( V \), the latency \( \delta(u, v) \) for sending a message between them via a TCP connection is assumed to be a constant non-negative value. This latency encompasses all transmission delays including in-network factors such as propagation and queuing. The value of \( \delta(u, v) \) is influenced by several variables, including the size of the transmitted messages, the internet access bandwidth available at nodes \( u \) and \( v \), the physical distance between these nodes, and the level of network congestion. These factors are considered to be slowly varying relative to the timescale of our algorithm. Additionally, each node \( v \) in \( V \) incurs a fixed processing time \( \Delta_v \) for handling membership update messages, which varies based on the computing performance of the node.

We posit that membership status is periodically broadcast across the network. When a node \( u \) either initiates a membership message or receives one from an incoming neighbor, it promptly relays this message to each of its outgoing neighbors. The completion of this relay for each neighbor \( v \) occurs within a designated time \( \delta(u,v) \). This model assumes that the broadcast mechanism is immediate and sequential, ensuring rapid dissemination of membership updates throughout the network.

At any given time, each node in the network maintains \( \log(N) \) outgoing connections (\( d_{out} \)) and an equivalent number of incoming connections (\( d_{in} \)). To maintain the membership protocol, each node also keeps a local database, which is routinely updated through message exchanges with neighboring nodes. Newly joined peers acquire the membership list from existing nodes. We assume that each node is aware of the IP addresses of all other nodes in the network.

\subsection{Performance Metrics}
Given a fully connected P2P network, our objective is to construct one or multiple rings to minimize the diameter. The degree-constrained minimum diameter problem is an NP-hard problem \cite{adriaens2022diameter}.

\section{DGRO: Optimize Ring Topology over graphs}\label{sec:DGRO}

\subsection{Degree Constrained Subgraph with Minimum Diameter }
Given a weighted complete graph $ G = (V, E) $ where $ V $ is the set of vertices and $ E $ is the set of edges with associated weights $ w: E \rightarrow \mathbb{R}^+ $, the problem is to find a subgraph $G'= (V, E')$, where: $V$ is the set of all vertices from $ G $, $E' \subseteq E $, and each vertex $ v \in V $ has degree not exceeds $ K $ in $ G' $. The objective is to minimize the diameter $ D(G') $ of the subgraph $ G' $. The diameter is defined as: \begin{equation}
   D(G') = \max_{u, v \in V} d(u, v), 
\end{equation}
where $ d(u, v) $ representing the shortest path distance between vertices $ u $ and $ v $ in $ G' $, considering the weights of the edges. This problem seeks to optimize the connectivity and compactness of $ G' $ by minimizing its diameter while maintaining the degree constraint $ K $ for each vertex.

\subsection{Ring Construction}
A solution is constructed by sequentially adding edges into a partial solution \cite{adriaens2022diameter}.
\begin{itemize}
    \item \textbf{Input}: a complete graph $G=(V, E)$, initial subgraph $G_0 = (V, E_0)$, weights matrix $W$ and degree constraint $K$, where $|V| = N, W\in \mathbb{R}^{+N\times N}, K \in \mathbb{Z}^+$.
    \item \textbf{Sequential Addition}: Add an edge $e_t$ to the partial solution $G_t$  $$G_t=(V, E_t) \xrightarrow{e_t} G_{t+1}=(V, E_t\cup\{e_t\}), $$
    and the degree constraint, $degree(v) \leq K, \forall v \in V$, is satisfied.
    
    \item \textbf{Selection Strategy}: Select the edge with the highest score based on a scoring function $F(G, G_t)$. $$e_t = \argmax_{e\in E \setminus E_t} F(G, G_t, e).$$ The \textit{nearest-neighbour heurisitic} adopts the edge weight $w(e)$ as the score function, namely $F(G, G_t, e) = w(e)$. 
    \item \textbf{Termination}: $|E_t| \geq K \cdot N$ or all left edges $e \in E \setminus E_t$ cannot satisfy the degree constraint\footnote{\shixun{e.g. for $N=20, K=4$, the front $19$ nodes consist a graph with degree $4$, while the node $20$ still has degree $2$ and no more edges can be connected between node $20$ to any nodes from $1-19$.}}.
\end{itemize}

Defining a scoring function is an implicit task without a straightforward functional expression. Using the edge weight as the scoring function is a useful heuristic, but it is not exactly aligned with the goal of minimum diameter. In contrast, a neural network can develop a more effective scoring function than human-designed heuristics. By training on rewards derived from sampling, the neural network can generate scores that better reflect the impact of a selection on future graph construction and the diameter. We next formulate the diameter optimization problem with Markov Decision Process (MDP).

\begin{figure}
    \centering
    \includegraphics[width=\linewidth]{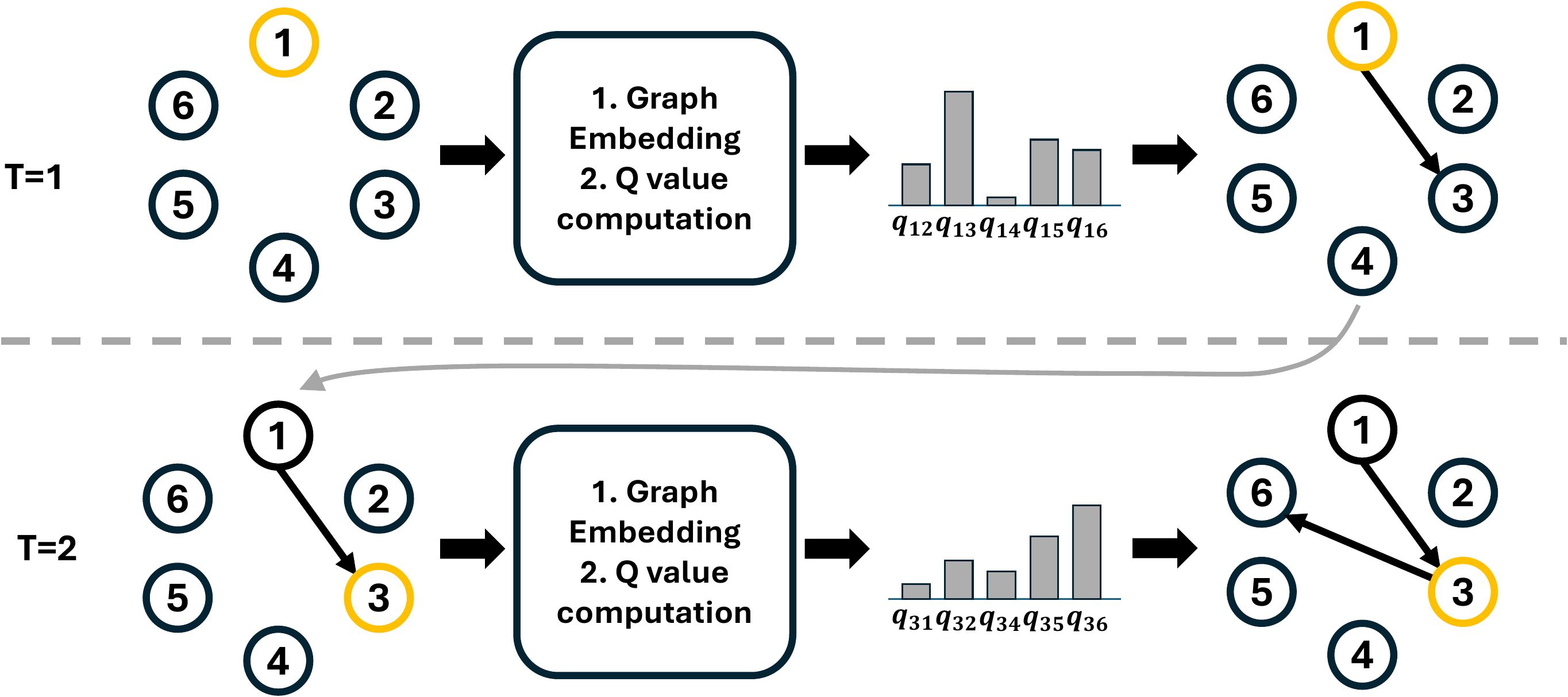}
    \caption{Node selection with DGRO}
    \label{fig:workflow}
    \vspace{-3mm}
\end{figure}
\subsection{Markov Decision Process Formulation}
In the formulation discussed last section, $e_t$ is chosen from $E \setminus E_t$, requires scoring $K\cdot N$ to $N^2$ edges, depend on $|E_t|$. To reduce complexity, we select $e_t$ from $\{(v_t, u)|u\in V\} \setminus E_t$, where $v_t$ is the end node of last chosen edge $e_{t-1}=(v_{t-1}, v_t)$. Hence, the state becomes $S_t = (W, W_t, v_t)$ and the score function is 
$$ F(G, G_t, e_t) = Q(S_t, u; \Theta).$$
$Q(S_t, u; \Theta)$ is a neural network takes $S_t$ as input and output a score of edge $(v_t, u)$, where $\Theta$ is the network weights. The action space is defined as \([1, \ldots, N]\), where \(N\) represents the number of nodes within the graph. The state is characterized by the latency matrix in conjunction with the topology that has been constructed up to the current step.  

The reward function is computed as the difference in diameters between consecutive states, specifically, \( r(S_t, S_{t+1} = D(G_t) - D(G_{t+1}) - \alpha w(a_t, a_{t + 1}) \), where $D$ is the diameter, \(S_t\) and \(S_{t+1}\) ($a_t$ and $a_{t+1}$) represent the state (action) at time \(t\) and \(t+1\),  respectively. With this reward function, the diameter term of cumulative reward will be $$\sum_{t = 0}^{T - 1} D(G_{t-1}) - D(G_{t}) = D(G_0) - D(G_{T}).$$ Because we start from an empty topology, $D(G_0)$ is set to a constant. Hence, the cumulative reward of an episode will become the diameter of the final topology, namely $D(G_{T})$, which is exactly our optimization objective. When the graph $G_t$ is not connected, the diameter of the largest connected component is adopted. Besides, we augment the reward function with an additional term $\alpha w(a_t, a_{t + 1})$ to mitigate the incidence of zero rewards because the diameter tends to stabilize and becomes difficult to reduce further once more than half of the edges have been added to the topology. \(w(a_t, a_{t+1})\) represents the latency of the connection between nodes \(a_t\) and \(a_{t+1}\), and \(\alpha\) is a coefficient controlling the weight of the latency term. This modification to the reward function is intended to incentivize actions that not only improve the diameter but also consider the latency of individual connections, thus enhancing the overall network performance.

\begin{algorithm}
\caption{Diameter-Guided Ring Construction.}
\begin{algorithmic}
\State \textbf{input:} start node $v_0$, Q-network $Q(:\Theta)$, latency matrix $A$
\State \textbf{output:} optimized ring ($v_0\rightarrow \cdots\rightarrow v_{N-1}\rightarrow v_0$). 
\For{step $t = 1$ to $T$}
\State $v_t = 
\arg\max_{v} \widehat{Q}(S_t, v; \Theta)$
\State Add $v_t$ to partial solution: $S_{t+1} := (S_t, v_t)$
\EndFor
\end{algorithmic}\label{alg:ring_construction}
\end{algorithm}

Algorithm \ref{alg:ring_construction} demonstrates the ring construction with Q value as a score function. In Figure \ref{fig:workflow}, we provide a specific example of our method in action. Starting from node 1, our algorithm calculates the Q-values \(q_{12}, q_{13}, \ldots, q_{16}\) for connections from node 1 to nodes 2 through 6. The algorithm selects the node associated with the highest Q-value among these, which in this example is node 3 due to \(q_{13}\) being the largest. Consequently, the action taken in this step is to establish a connection from node 1 to node 3. In the subsequent step, the algorithm recalculates the Q-values for connections from node 3 to the remaining nodes and selects the next action in a similar manner, continuously optimizing the network structure based on these Q-value assessments. This sequential decision-making process effectively builds the network topology by linking nodes that maximize the predicted Q-value, thereby optimizing the overall connectivity based on the learned Q-function.

\subsection{Graph Embedding}
Due to the substantial memory demands and learning inefficiencies of using the adjacency matrix as a direct input, which would require at least \(N^2\) network parameters, we've opted for a graph embedding approach. This method encodes the neighborhood and latency information of each node into a one-dimensional vector with a feature dimension \(p\). This embedding technique effectively reduces the complexity and enhances the scalability of our model, making it more feasible for handling large networks without a significant loss in performance or detail.

 Eqn. (\ref{eqn:embed1}) is the embedding strategy for the complete graph $G$ and the partial solution subgraph $G_t$.  In Eqn. (\ref{eqn:embed1}), $\mu_v^{(t)} \in \mathbb{R}^p$ is the embedding vector of node $v$ after $t$ iteration, the initial embedding is a zero vector, $x_v$ is the degree of node $v$, $\mathcal{N}(v)$ is the neighbour of node $v$, $\theta_1 \in \mathbb{R}, \theta_2, \theta_3 \in \mathbb{R}^{p\times p}, \theta_4 \in \mathbb{R}^{p}$. By applying Eqn. (\ref{eqn:embed1}) for $T$ iterations, we get the final embedding vector $\mu_v^{(T)}$. After we get the node embedding $\mu_v^{(T)}$, we combine the embeddings of all nodes sum, source node $v_{t+1}$, target node $u$ and the weight $w(v_{t+1}, u)$ into $x \in \mathbb{R}^{3p+1}$, as shown in Eqn. (\ref{eqn:feature}). Finally, $x$ is forwarded to a MLP to get the final score of selection edge $e_{t+1} = (v_{t}, u)$.

\begin{equation}
\begin{split}
\mu_v^{(t+1)} \leftarrow \operatorname{relu} &\left( \theta_1 x_v+\theta_2 \sum_{u \in \mathcal{N}(v)} \mu_u^{(t)} \right. \\
&\left. +\theta_3 \sum_{u \in \mathcal{N}(v)} \operatorname{relu}\left(\theta_4 w(v, u)\right) \right)
\end{split}
\label{eqn:embed1}
\end{equation}
\begin{equation}
\begin{split}
    x\leftarrow [w(v_t, u), &\theta_5 \sum_{v \in V} \mu_u^{(T)}, \theta_{6} \mu_{v_t}^{(T)}, \theta_{7} \mu_u^{(T)}
    % \\ 
    % &\theta_{8} \sum_{v \in V} \nu_v^{(T)}, \theta_{9} \nu_{v_t}^{(T)}, \theta_{10} \nu_u^{(T)} 
    ]
\end{split}
\label{eqn:feature}
\end{equation}
\begin{equation}
    \widehat{Q}(S_t, u; \Theta)= \theta_{10}^\top \operatorname{relu}\left( \theta_{9} \operatorname{relu} \left(\theta_{8} \operatorname{relu}\left( x\right)\right)\right)
    \label{eqn:Q_net}
\end{equation}

\begin{figure}[t]
    \centering
    \includegraphics[width=\linewidth]{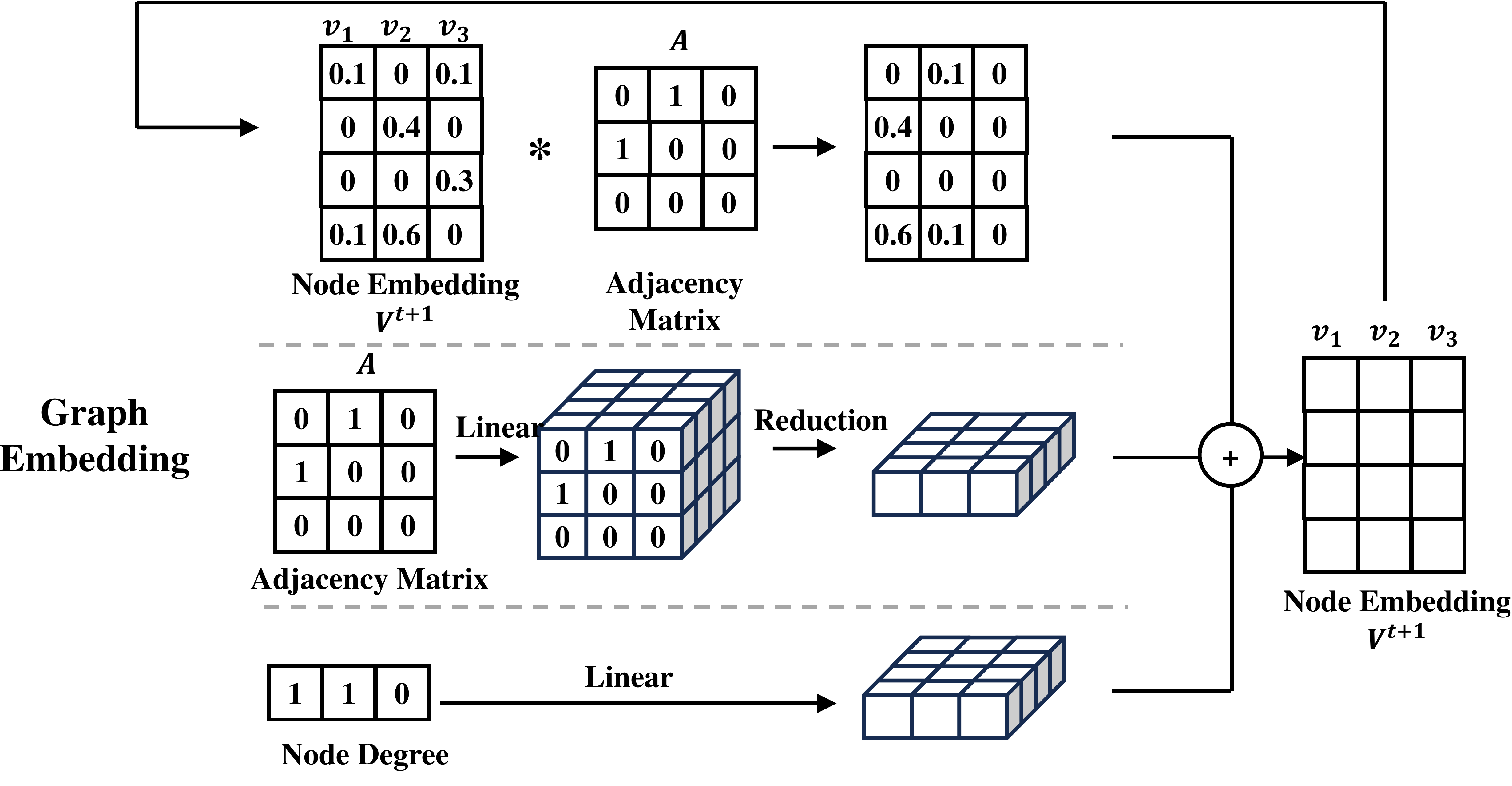}
    \caption{Graph Embedding}
    \label{fig:graph_embedding}
    \vspace{-5mm}
\end{figure}

Figure \ref{fig:graph_embedding} illustrates how we optimize performance by transforming embeddings into matrix multiplication operations. The first row depicts the multiplication of node embeddings with the latency matrix, which corresponds to the term \(\theta_2 \sum_{u \in \mathcal{N}(v)} \mu_u^{(t)} \) in Equation \ref{eqn:embed1}. The second row processes the latency matrix through a linear layer first, then performs a weighted reduction along the column direction based on the weights of the latency matrix, aligning with the term $\theta_3 \sum_{u \in \mathcal{N}(v)} \operatorname{relu}\left(\theta_4 w(v, u)\right.$ 

% \begin{equation}
% \begin{split}
% \nu_v^{(t+1)} \leftarrow \operatorname{relu} & \left( \theta_5 x_v+\theta_6 \sum_{u \in \mathcal{N}(v)} \nu_u^{(t)} \right. \\
% & \left. +\theta_7 \sum_{u \in \mathcal{N}(v)} \operatorname{relu}\left(\theta_8 w(v, u)\right)\right)
% \end{split}
% \label{eqn:embed2}
% \end{equation}

\subsection{Learning Algorithm}

Standard 1-step \textit{Q}-learning updates the parameters of the function approximator at each step within an episode by performing a gradient descent step aimed at minimizing the squared loss.
\begin{equation}
    \left(y-\widehat{Q}\left(S_t, v_t; \Theta\right)\right)^2,
\end{equation}
where $y = r(S_t, S_{t+1}) + \gamma\cdot\widehat{Q}\left(S_{t+1}, v_{t+1}; \Theta\right)$. 

Algorithm \ref{alg:q_learning} illustrates the training framework with an experience replay mechanism. Initially, a memory buffer is established to store transitions, aiding in the machine-learning process across multiple epochs. Each epoch involves sampling a graph and constructing a solution iteratively, where node selections are based either on randomness (to explore diverse paths) or on maximizing expected rewards via a Q-value function (to exploit known strategies). Critical to this approach is the dynamic updating of the solution state with each new node addition, influenced by prior actions stored in the replay memory. These experiences are randomly revisited to update the learning model's parameters through stochastic gradient descent, ensuring continual refinement of the strategy. The process iteratively enhances the model's ability to predict and execute optimal actions, culminating in a robust set of parameters that define the most effective decision-making process for the types of graphs encountered.

\begin{algorithm}
\caption{Q-learning for DGRO}
\begin{algorithmic}
    \State Initialize experience replay memory $\mathcal{M}$ to capacity $N$
    \For{epoch $e = 1$ to $L$}
        \State Draw graph $G$ from distribution $\mathcal{D}$
        \State Initialize the state to empty $S_1 = ()$
        \For{step $t = 1$ to $T$}
            \State $v_t = \begin{cases} 
            \text{random node } v \in S_t, & \text{with probability } \epsilon \\ 
            \arg\max_{v} \widehat{Q}(S_t, v; \Theta), & \text{otherwise}
            \end{cases}$
            \State Add $v_t$ to partial solution: $S_{t+1} := (S_t, v_t)$
            \If{$t \geq n$}
                \State Add tuple $(S_{t-1}, v_{t-1}, R_{t-1,t}, S_t)$ to $\mathcal{M}$
            \EndIf
            \State Sample random batch from $B \sim \mathcal{M}$
            \State Update $\Theta$ by SGD over $B$
        \EndFor
    \EndFor
    \State \Return $\Theta$
\end{algorithmic}\label{alg:q_learning}
\end{algorithm}

% \subsection{Diameter Estimation}

\section{Self Adaptive Ring Topology Optimizations}\label{section:ring_topology_matters}
In subsequent experiments, deep Q-learning proved capable of outperforming heuristic-based solutions for networks with up to 200 nodes. However, as the number of nodes increases, the computational power and training time required also increase significantly, limiting the scalability of Q-learning. This is particularly evident in K-ring topologies, where the number of required steps scales with \( K \times N \), and \( K \) typically takes a value of \( \log_2(N) \). For instance, with \( N = 500 \), around 4500 steps are necessary. Given these constraints, simple rule-based heuristic algorithms remain essential under such conditions. 

Our findings indicate that the choice of ring based on different heuristics significantly impacts the total diameter of peer-to-peer (P2P) topologies. Therefore, we have designed a decentralized scheme that allows the P2P network to automatically adjust its ring topology based on the current latency conditions, enhancing network performance dynamically. This approach facilitates more efficient network configurations that adapt in real-time to changing operational environments.

\begin{algorithm}
\caption{Gossip-based Latency Measurement}
\begin{algorithmic}[1]
\State \textbf{Input:} network $G=(V, E)$, time period $T$, \#samples $K$.
\State \textbf{Output:} $\overline{L_{local}}$, $\overline{L_{global}}$, $\overline{L_{min}}$

\State Initialize $L_{local} \gets 0$, $L_{global} \gets 0$, $L_{min} \gets 0$

\For{each node $u \in G$}
    \State Randomly select $\{r_i\}_{i=1}^K$, s.t. $(u,v_i)\in E$.
    \State Randomly select $\{v_i\}_{i=1}^K$.
    \State $L_{local}=\frac{1}{K}\sum_{i=1}^K L(u,r_i)$
    \State $L_{global}\gets \frac{1}{K}\sum_{i=1}^kL(u,v_i)$
    \State $L_{min}\gets \min_{i=1}^kL(u,v_i)$
\EndFor
\State Initialize zero to $L_{local}, L_{global}$, $L_{min}$, $M$
\For{each node $u \in G$}
    \For{each round}
        \State Gossip $(W_{local}, L_{global}, L_{min})$ to neighbors
        \State Accumulate received $L_{local}$, $L_{global}$, and $L_{min}$
        \State Update message count $M \gets M + 1$
    \EndFor
    \State Wait for $T$ for convergence
\EndFor

\State Compute averages: 
    \State $\overline{L_{local}} \gets \sum L_{local} / M$
    \State $\overline{L_{global}} \gets \sum L_{global} / M$
    \State $\overline{L_{min}} \gets \sum L_{min} / M$\\
\Return $\overline{L_{local}}, \overline{L_{global}}, \overline{L_{min}}$   
% \If{$\frac{\overline{L_{local}} - \overline{L_{min}}}{\overline{L_{avg}} - \overline{L_{min}}} < \epsilon \text{ or } \frac{\overline{L_{local}} - \overline{L_{min}}}{\overline{L_{avg}} - \overline{L_{min}}} > 1 - \epsilon$}
%     \State Network is either too cluttered or too random
% \Else
%     \State Network diameter is balanced
% \EndIf
\end{algorithmic}\label{alg:heuristic_selection}

\end{algorithm}
Algorithm \ref{alg:heuristic_selection} assesses network topology based on latency characteristics. In Algorithm 3, each node randomly samples \( K \) nodes from both its existing connections and the entire network to test latencies, denoted as \( L_{local} \) and \( L_{global} \), respectively. The algorithm proceeds to calculate the average values of both \( L_{local} \) and \( L_{global} \), as well as the minimum value within \( L_{global} \). Using a gossip protocol, these three values from all nodes are then aggregated and averaged. The process concludes by summing up and outputting these aggregated measurements, effectively providing a comprehensive view of the network's latency landscape. This method allows for a decentralized assessment of network performance, facilitating optimal adjustments to the network configuration based on real-time latency data. 

Once the period \( T \) has elapsed, each node calculates the ratio $\rho = \frac{\overline{L_{local}} - \overline{L_{min}}}{\overline{L_{avg}} - \overline{L_{min}}}$ to assess whether the current topology is either too concentrated or too random. This metric provides a standardized measure of network structure by comparing the dispersion of local latencies relative to the minimal and average latencies observed across the network. A value close to 0 or 1 indicates a topology that is either too clustered or excessively dispersed, respectively, suggesting potential inefficiencies in data routing and network resilience. This evaluation helps in identifying areas where the network topology may need adjustments to optimize performance and reliability. 

To manage the network topology efficiently based on latency distribution, we can set a threshold \( \epsilon \). If the calculated ratio $\rho > \epsilon$, it indicates that the distribution is too clustered. Then an additional pre-designed random ring can be introduced to diversify node connections and decrease network clustering. Conversely, if the distribution appears too sparse, with \( \rho > 1 - \epsilon \), incorporating the shortest ring can help tighten the network's topology by connecting nodes that are closer together in terms of latency. This adaptive approach allows for dynamic adjustments to the network's structure, optimizing performance based on real-time analysis of latency characteristics. The shortest ring is constructed by sequentially selecting the nearest available neighbor.

\subsection{Impact of different ring topology on p2p overlays}
\begin{figure}[htp]
    \centering
    \includegraphics[width=0.9\linewidth]{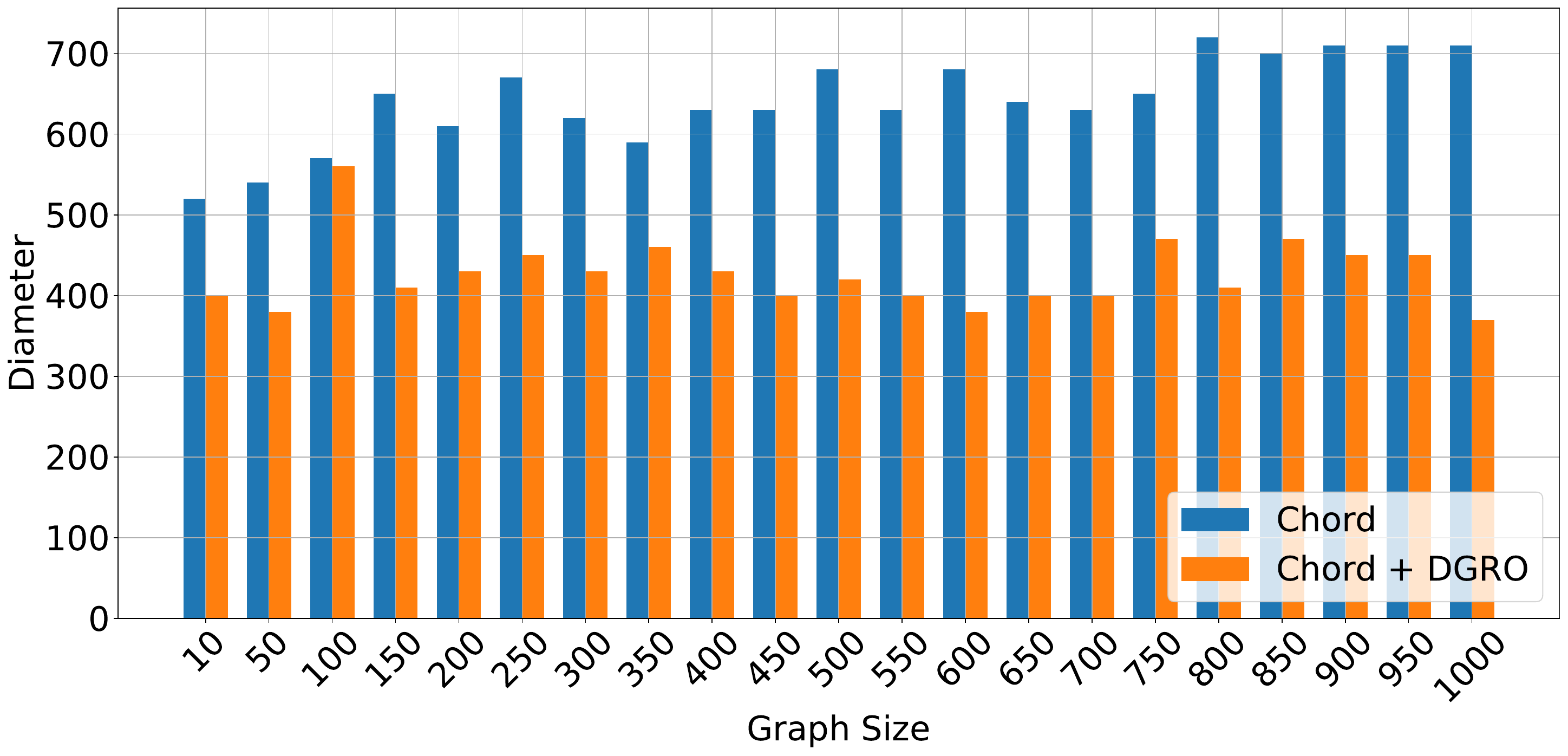}
    \caption{DGRO helps Chord reduce diameters.}
    \label{fig:chord_random_vs_nn}
     \vspace{-3mm}
\end{figure}
\subsubsection{Chord}
Chord \cite{stoica2001chord} assigns each node and data item a unique identifier, allowing nodes to locate resources within a logarithmic number of steps relative to the total number of nodes in the network. The unique identifier is given by a hashing function, which forms a logical ring among all nodes. Due to the randomness of the logical ring, Chord shows a $\rho$ close to 1. By replacing the random ring with the shortest ring, the diameter is reduced by 10\% to 40\%, as shown in Figure \ref{fig:chord_random_vs_nn}.

\begin{figure}[htp]
   \centering
\includegraphics[width=0.9\linewidth]{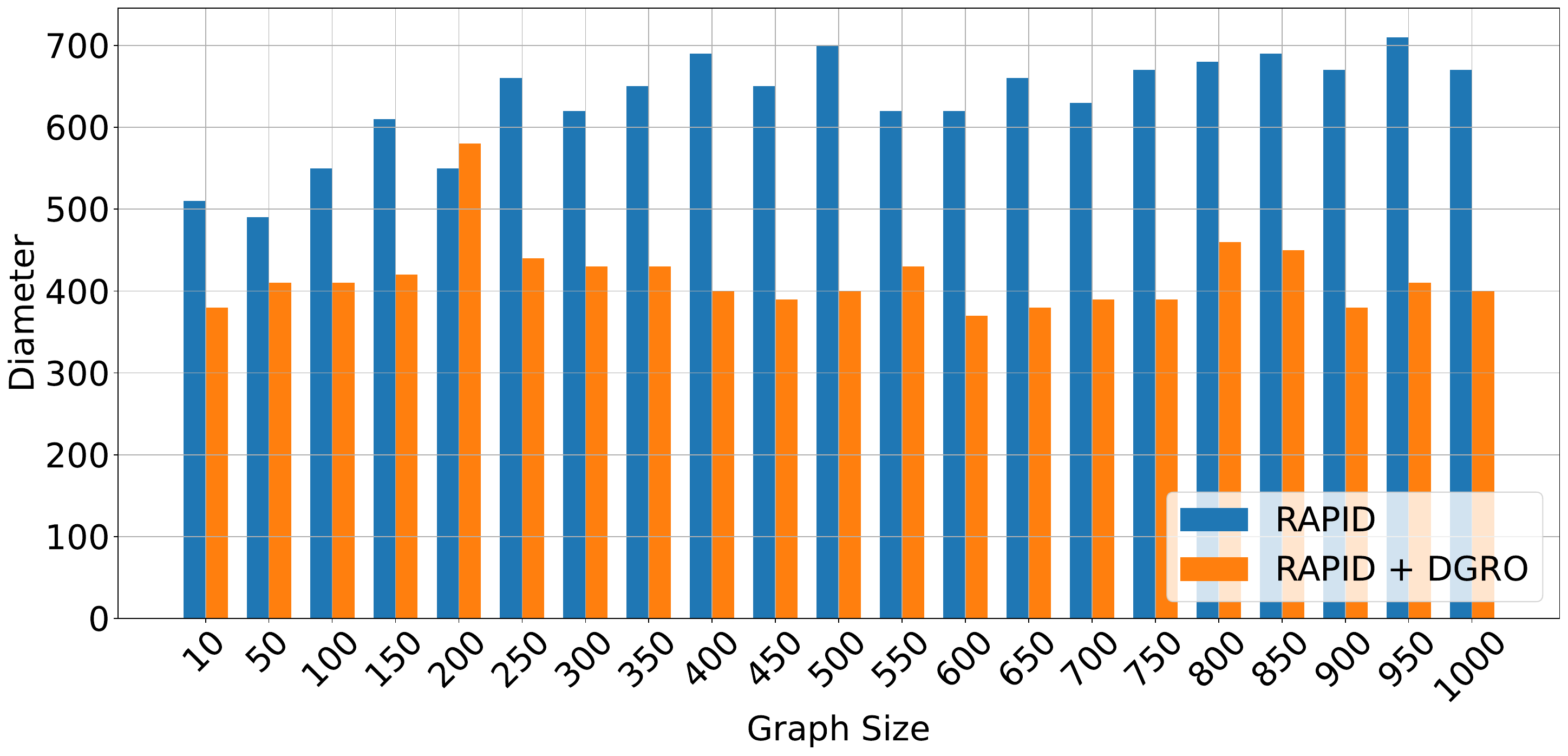}
    \caption{DGRO helps RAPID reduce diameters.}
    \label{fig:RAPID_random_vs_nn}
    \vspace{-3mm}
\end{figure}

\subsubsection{RAPID}
 Although the K-Ring topology \cite{suresh2018stable} is a good expander, the system model does not consider the latency between each node. The consistency hash function used to generate the K-Ring topology is fully randomized, which results in a high diameter, as shown in Figure \ref{fig:Perigee_random_vs_nn}. By switching one of the random rings to the shortest ring, the diameter get a significant decrement up to 43\%. 

\subsubsection{Perigee}
\begin{figure}[htp]
    \centering
    \includegraphics[width=0.9\linewidth]{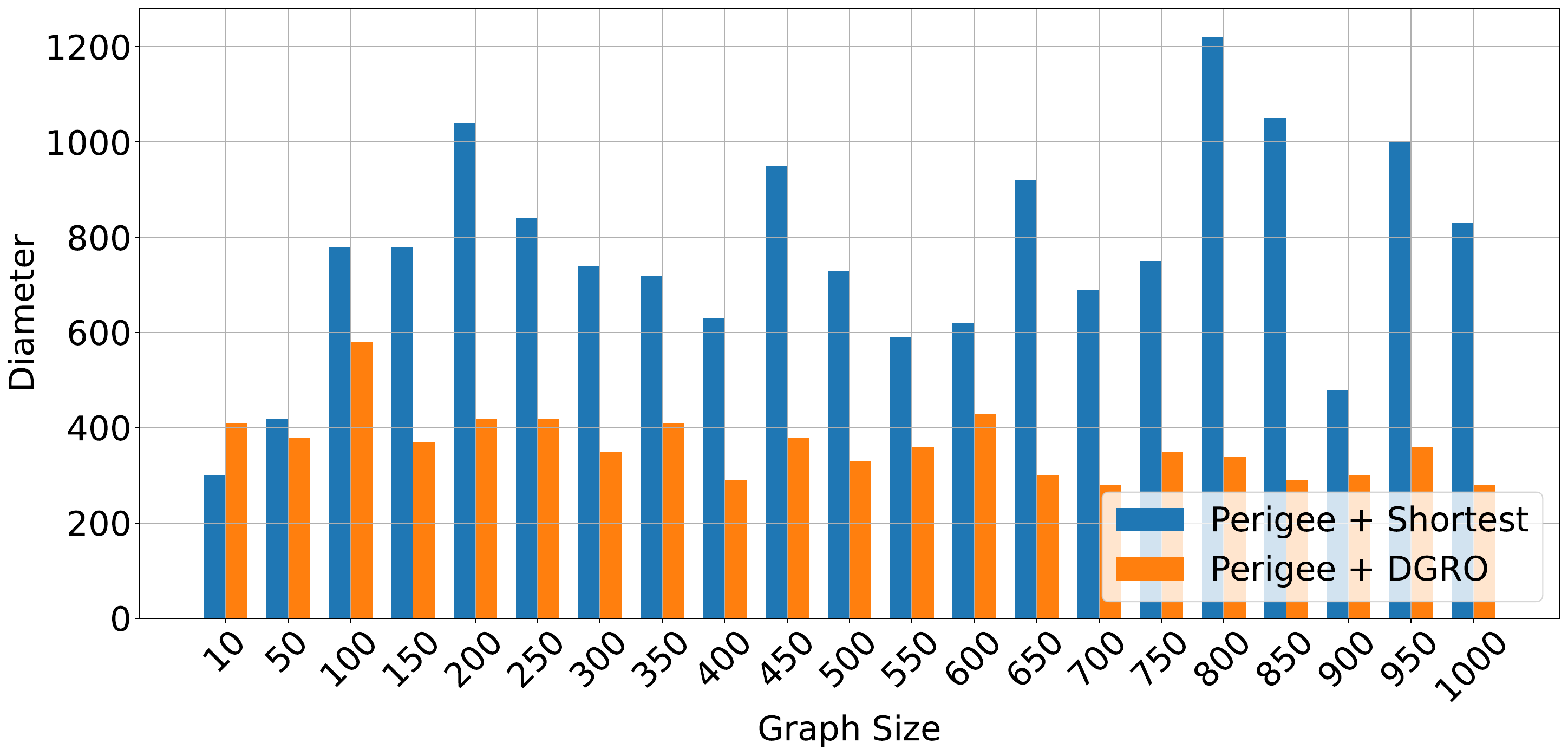}
    \caption{DGRO finds better diameters for Perigee.}
    \label{fig:Perigee_random_vs_nn}
\end{figure}

 Perigee \cite{mao2020perigee} is a nearest-neighbor-based method and typically does not require a specific ring. Each node organizes its neighbors based on the timestamp of receiving a random global broadcast sourced from a neighboring node. Figure \ref{fig:Perigee_random_vs_nn} compares the diameter of Perigee after adding a random ring and the shortest ring. Random ring has a low diameter compared to the shortest ring. For a high network size close to 1000, the improvement in diameter is up to 200\%.

\section{Parallel}\label{sec:parallel}
In addition to our existing strategies, we also experimented with a parallel construction approach for forming rings to decrease the number of sequential steps required in ring construction. This method is aimed at enhancing scalability by distributing the workload across multiple nodes simultaneously, thus significantly speeding up the overall process of ring formation.

This method involves dividing the \( N \) nodes into \( M \) partitions, thereby transforming a process that would typically require \( N \) sequential steps into a parallel procedure across \( M \) partitions. Each partition concurrently executes \( \frac{N}{M} \) sequential steps internally. This approach effectively reduces the total time required for ring construction by parallelizing the work, thus enhancing scalability and efficiency in handling larger datasets or networks.

\begin{figure}
    \centering
    \includegraphics[width=\linewidth]{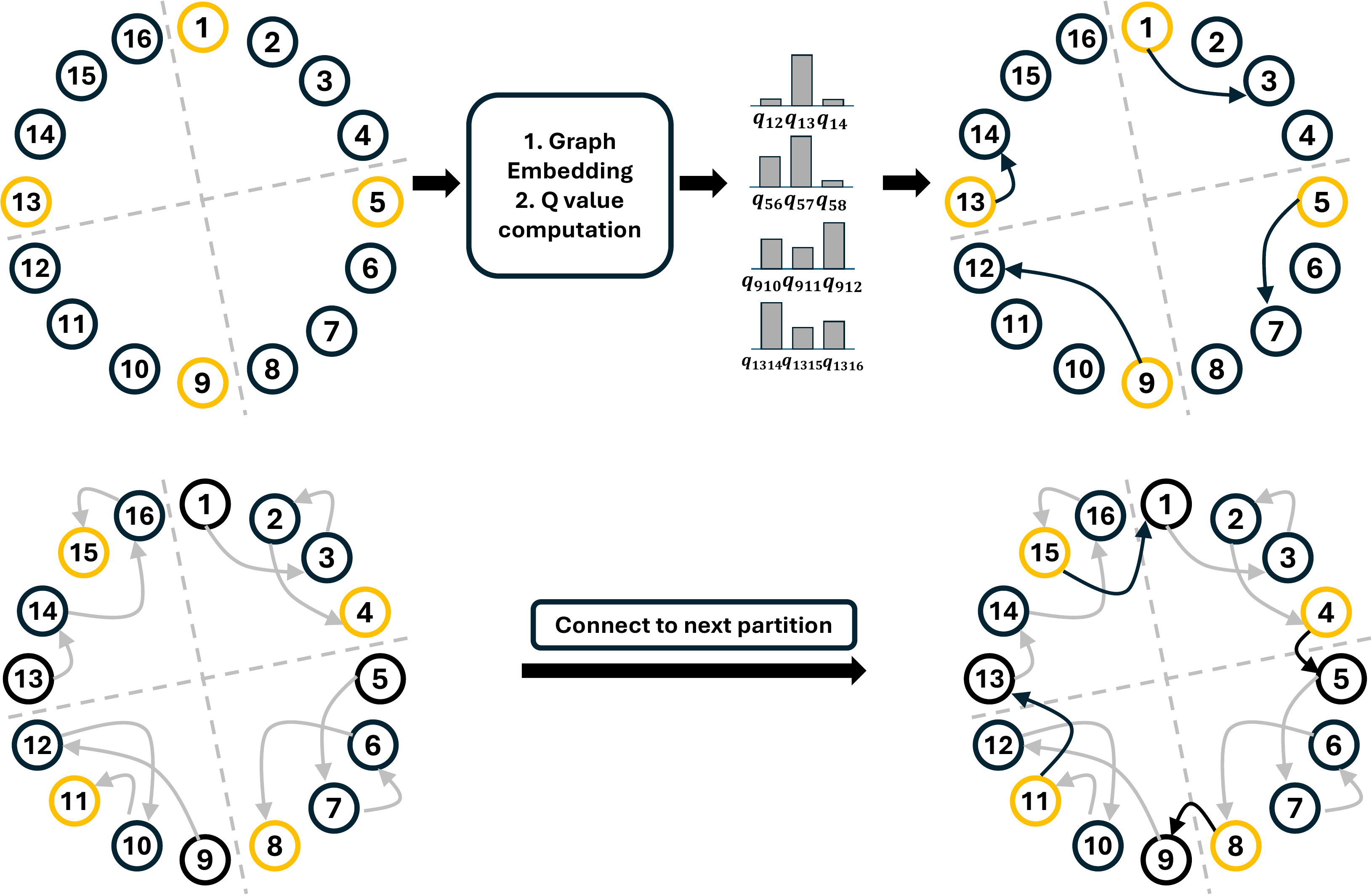}
    \caption{Parallel Ring Construction Workflow}
    \label{fig:parallel_workflow}
    \vspace{-5mm}
\end{figure}

% \begin{algorithm}
%     \caption{Parallel Diameter-Guided Ring Construction.}
%     \begin{algorithmic}
%     \State \textbf{input:} start node $v_0$, Q-network $Q(:\Theta)$, initial state $S_0$, parallel-level $P$
%     \State \textbf{output:} optimized ring ($v_0\rightarrow \cdots\rightarrow v_{N-1}\rightarrow v_0$). 
%     \State BlockID
%     \For{step $t = 1$ to $T$}
%     \State $v_t = 
%     \arg\max_{v} \widehat{Q}(S_t, v; \Theta)$
%     \State Add $v_t$ to partial solution: $S_{t+1} := (S_t, v_t)$
% \EndFor
% \end{algorithmic}
% \end{algorithm}

\begin{algorithm}
\caption{Parallel Ring Construction}
\begin{algorithmic}[1]
\State \textbf{Input:} Total nodes $N$, Number of partitions $M$
\State \textbf{Output:} Constructed ring across partitions

\State Initialize an empty structure for the ring
\State Calculate partition size: $partition\_size = N / M$
\State Divide nodes into $M$ partitions

\For{each partition $P_i$ in parallel}
    \State Initialize local ring for $P_i$
    \For{step $= 1$ to $partition\_size$}
        \State Select the next node in $P_i$
        \State Add node to the local ring of $P_i$
        \If{step $<$ partition\_size}
            \State Node selection with DGRO
        \Else
            \State Connect last node in $P_i$ to first node in $P_{i+1}$
        \EndIf
    \EndFor
    \State Merge local ring of $P_i$ into the main ring structure
\EndFor

\State If there are any nodes left (due to integer division), add them sequentially to the main ring
\State \Return the main ring
\end{algorithmic}\label{alg:parallel}
\end{algorithm}
Algorithm \ref{alg:parallel} initiates by setting up a structure to accommodate the entire ring and then splits the total nodes \(N\) into \(M\) partitions, with each partition comprising roughly \(N/M\) nodes. In the parallel execution phase, each partition \(P_i\) independently constructs its segment of the ring. This is done by sequentially connecting nodes within the partition until all are linked, effectively closing the loop with the last node connecting back to the first in each partition. Once all partitions have completed their individual segments, these are merged into the main ring structure. To ensure the inclusivity of all nodes, especially when \(N\) is not perfectly divisible by \(M\), the remaining nodes are added sequentially to complete the ring.
% ### Explanation:
% - **Step 1-3**: The algorithm begins by initializing a structure to hold the complete ring and then divides the total number of nodes \( N \) into \( M \) partitions, each containing approximately \( N/M \) nodes.
% - **Step 4-12 (Parallel Execution)**: Each partition \( P_i \) constructs its segment of the ring in parallel. Nodes within each partition are sequentially connected until all nodes in the partition are linked, closing the loop in the final step of each partition.
% - **Step 13-14**: After each partition has completed its internal ring, these segments are merged into the main ring structure. Any nodes not perfectly divisible by \( M \) are added sequentially to ensure all nodes are included in the final ring.

% This pseudocode illustrates how dividing the workload into partitions can significantly accelerate the process of ring construction by leveraging parallel processing, which is particularly beneficial for large-scale systems.
\section{Performance Evaluation}
\label{sec:results}

We evaluate the performance of DGRO, and compare it against the baseline algorithms of $\S$\ref{section:ring_topology_matters}. Our experiments are based on a Python
simulator we built following the network model of $\S$\ref{section:system_model}. We describe the experimental setting in $\S$\ref{experimental_settings}. Following this, we evaluate Perigee
on a variety of different network conditions (\S\ref{sec:synthetic}–\S\ref{sec:realisitic}). All the mentioned \textit{shortest rings} represent rings constructed by nearest neighbors. 

\subsection{Experimental Settings}
\label{experimental_settings}
\subsubsection{Network settings} 
In our experiments, we evaluated the DGRO protocol using four distinct latency distributions to understand its performance under various network conditions. We used two synthetic distributions: a uniform distribution randomly sampled from the set $\{1, 2, \ldots, 10\}$, namely $X \sim \text{Uniform}(1, 10)$
and a Gaussian distribution with a mean of 5 and a standard deviation of 1: $Y \sim \mathcal{N}(5, 1^2)$. Additionally, we incorporated two realistic latency distributions, FABIRC and Bitnode. FABRIC is based on latency data collected from 17 physical sites, which include 14 sites across the US, one in Japan, and two in Europe. Bitnode utilized latency of 1000 nodes. The nodes are randomly sampled from a list of 9,408 nodes. These nodes are distributed across seven geographic regions: North America, South America, Europe, Asia, Africa, China, and Oceania. For our experiment, the network size, link propagation, processing, and transmission delays were set as follows: (1)\textit{Propagation Delay}: For FABRIC, we used the average one-hour measurement of one-way latency between each node pair.  The propagation latency between a node \( u \) at site \( i \) and a node \( v \) at site \( j \) is calculated as \( \text{latency}(i, j) + \text{latency}(u) + \text{latency}(v) \). The $\text{latency}(i, j)$ is measured from the FABRIC monitoring tool. For Bitnode, propagation latency was determined by geographical locations using the iPlane latency measurement dataset \cite{mao2020perigee}. (2) \textit{Message Size}: We assumed messages were small relative to the available bandwidth at the nodes, making the link propagation delays the dominant factor in message broadcasting delay. (3) \textit{Message Processing Time}: Each node was assigned a mean message processing time of $1$ ms. Additionally, each node was configured to create and accept connections logarithmic to the network size, specifically $\log(N)$ outgoing and up to $\log(N)$ incoming connections. This experimental setup allowed us to explore the effects of network scalability in sections §5.2 and §5.3, varying network sizes to assess performance under different conditions.
\subsubsection{Algorithm compared} We employed the topologies in CHORD, RAPID, and a nearest neighbor topology based on Perigee as compared baseline algorithm. Additionally, to establish a benchmark for the lowest possible network diameter, we utilized a genetic algorithm. For each graph instance, the genetic algorithm will search $100, 000$ topologies.

\subsubsection{Performance Metrics} To evaluate the effectiveness of each network topology, we calculate the network diameter using the NetworkX library. We conducted our tests across various network sizes, specifically $[50, 100, 150, ..., 1000]$. For the FABRIC network, we selected $17$ sites, assuming each site generates a varying number of nodes ranging from $1$ to $58$, resulting in total node counts from 17 to 986. The individual latencies \( \text{latency}(u) \) and \( \text{latency}(v) \) are assumed to follow a normal distribution with a mean of 5 and a standard deviation of 1, reflecting variability in node response times within each site. This setup allows us to model and analyze the impact of both inter-site and intra-site latencies on the overall network performance. For each network size, we performed 10 independent runs, each with randomly sampled link latencies. The results, showcasing the diameter for each network size, were plotted to provide a visual representation of performance across different scales. This methodological approach allows for a thorough analysis of how each topology performs under varying conditions and network sizes.

\subsection{Optimize Ring with Deep Q-learning}
We compare DGRO with Genetic Algorithm and random solution. The result of the genetic algorithm is searched over $100, 000$ graphs.
\begin{figure}[thp]
    \vspace{-3mm}
    \centering
    \includegraphics[width=\linewidth]{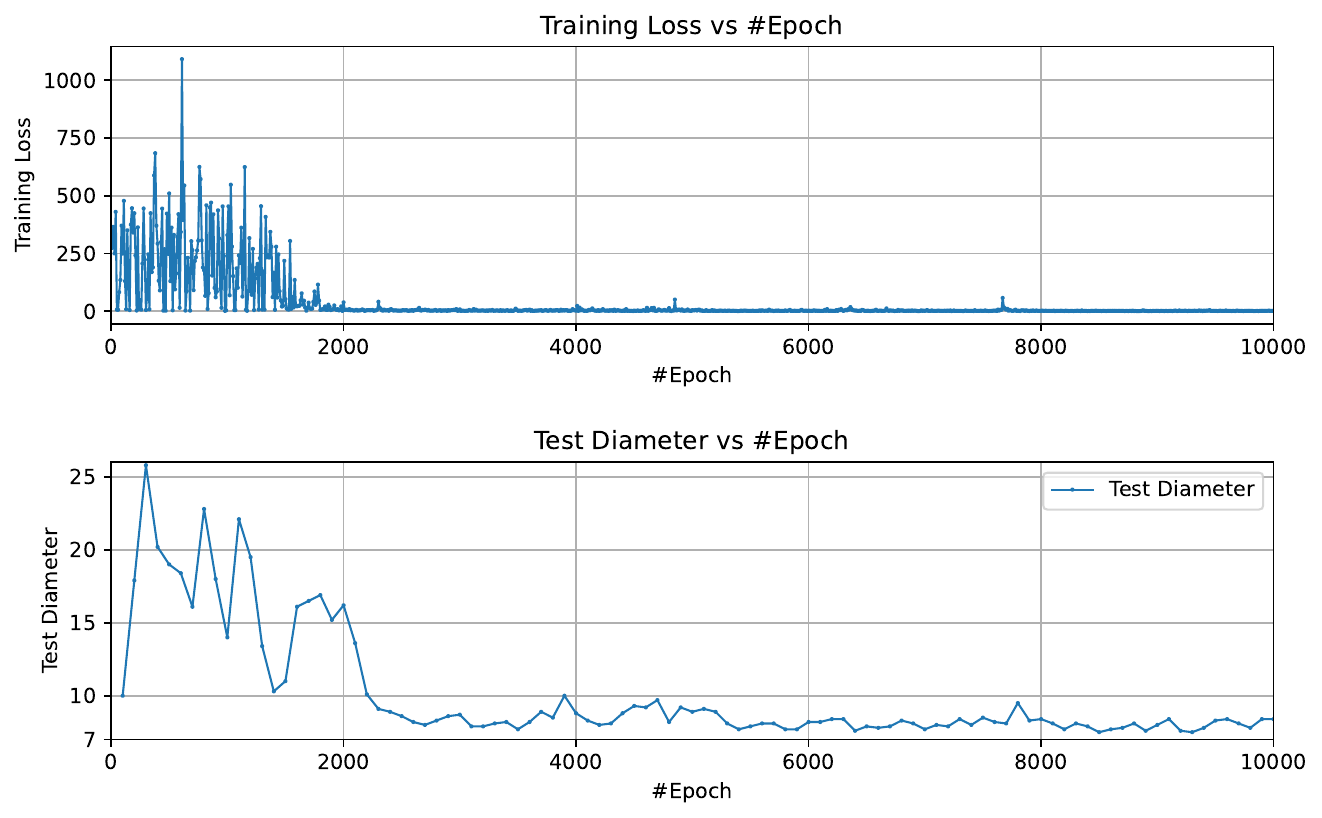}
    \caption{Training and Test Curve of DGRO.}
    \label{fig:training_curve}
    \vspace{-3mm}
\end{figure}
\subsubsection{Hyperparameter and training settings}
We configured the graph embedding with a feature dimension \( d = 16 \) and a learning rate \( \alpha = 5 \times 10^{-4} \). During training, we implemented an epsilon-greedy strategy where \( \epsilon \) is dynamically adjusted with the formula \( \epsilon = \max(1 - \frac{\text{epoch}}{2000}, 0.05) \). The replay buffer was set to a length of \( 10^6 \). We trained the model over \( 10^4 \) epochs with a batch size of 32. Each episode generates a new \( N \times N \) matrix, with each element uniformly and randomly selected from the set \([1, 2, \ldots, 10]\). Test graphs were generated using the same random sampling method.

As illustrated in Figure \ref{fig:training_curve}, the training curve converges after $2,000$ epochs, while the test curve stabilizes at a diameter of $10$ after $3,000$ epochs. This demonstrates the efficacy of the training process over extended periods, highlighting the model's ability to consistently achieve optimal network topology as training progresses.

\subsubsection{Performance Evaluation}
\begin{figure}[thp]
    \vspace{-3mm}
    \centering
    \includegraphics[width=\linewidth]{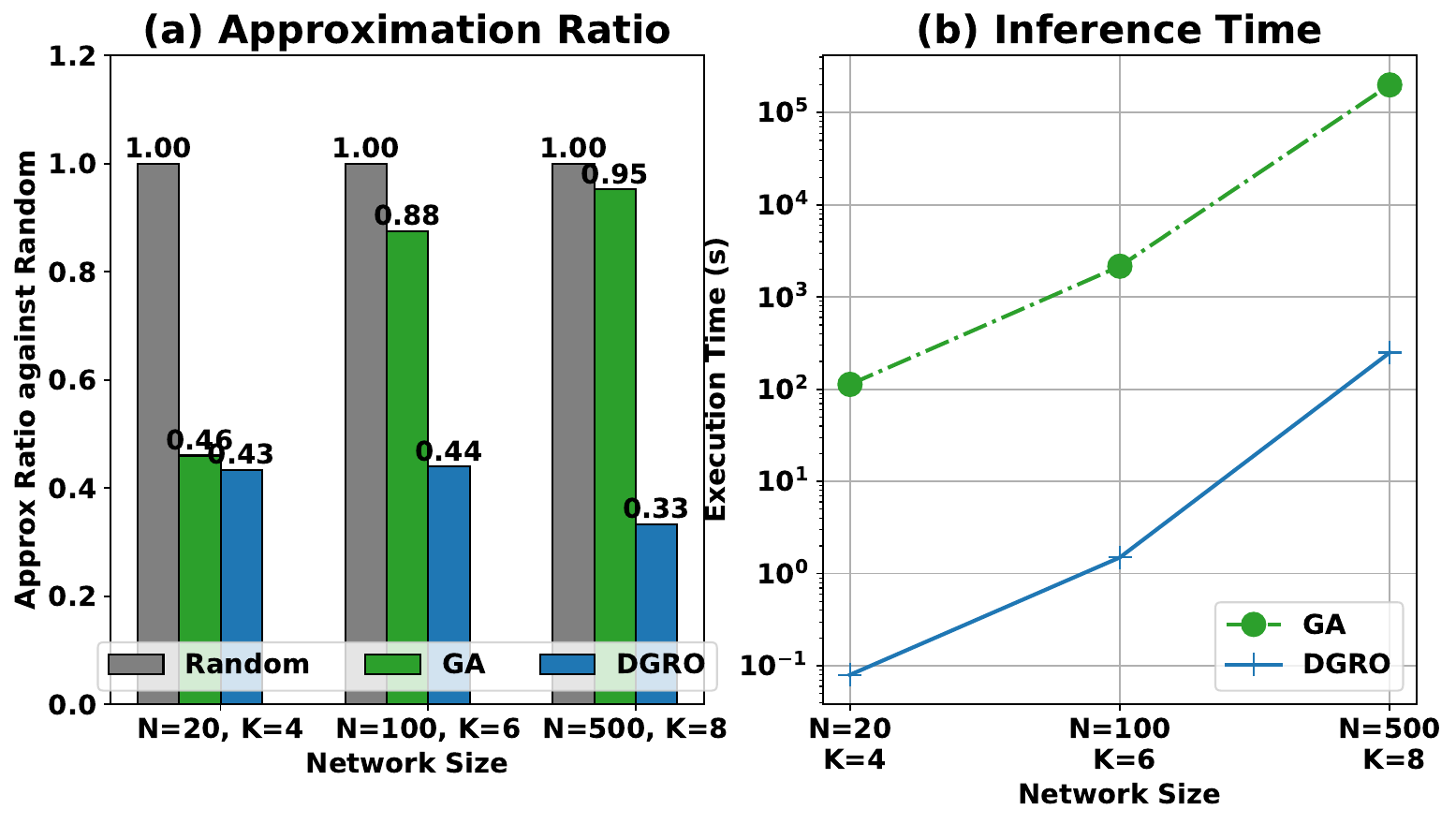}
    \caption{DGRO vs. Genetic Algorithm. The diameter is normalized by the result of a random ring.}
    \label{fig:dqn_diameter}
        \vspace{-3mm}
\end{figure}

We compare the diameter of the topologies constructed by DGRO with those generated by a genetic algorithm (GA) and a completely random ring. Figure \ref{fig:dqn_diameter} demonstrates the diameter while Figure \ref{fig:dqn_diameter} compares the inference time. Each method constructs a topology consisting of \( K \) rings, where each ring connects all nodes of the graph. For each graph instance, DGRO generates $10$ different \( K \)-ring topologies. The $10$ topologies are constructed with $10$ different starting nodes. Then we select the one with the best diameter. In contrast, the GA searches through \( 1 \times 10^5 \) different \( K \)-ring topologies to retain the one with the optimal diameter. For normalization, all calculated diameters are divided by the diameter of a random \( K \)-ring. This approach provides a clear comparison of the efficiency of DGRO against other methods in minimizing the network diameter across multiple configurations.

As shown in Figure \ref{fig:dqn_diameter} (a), DGRO outperforms the results of a brute force approach that iterates $100,000$ times, and it does so with considerably lower inference time. As the network size increases, the effectiveness of the genetic algorithm degrades to the random method due to the exponential growth in combinatorial possibilities. Meanwhile, DGRO consistently reduces the random diameter to $40\%$ of its original value. Figure \ref{fig:dqn_diameter} (b) shows DGRO has good scalability when problem size increases.

% \begin{figure}[htp]
%     \centering
%     \includegraphics[width=\linewidth]{figs/DQN_inference_time.pdf}
%     \caption{Caption}
%     \label{fig:dqn_inference_time}
% \end{figure}

\subsection{Synthetic Latency}
\label{sec:synthetic}
\subsubsection{Single heuristic ring}
\begin{figure}[htp]
\vspace{-4mm}
    \centering
    \includegraphics[width=\linewidth]{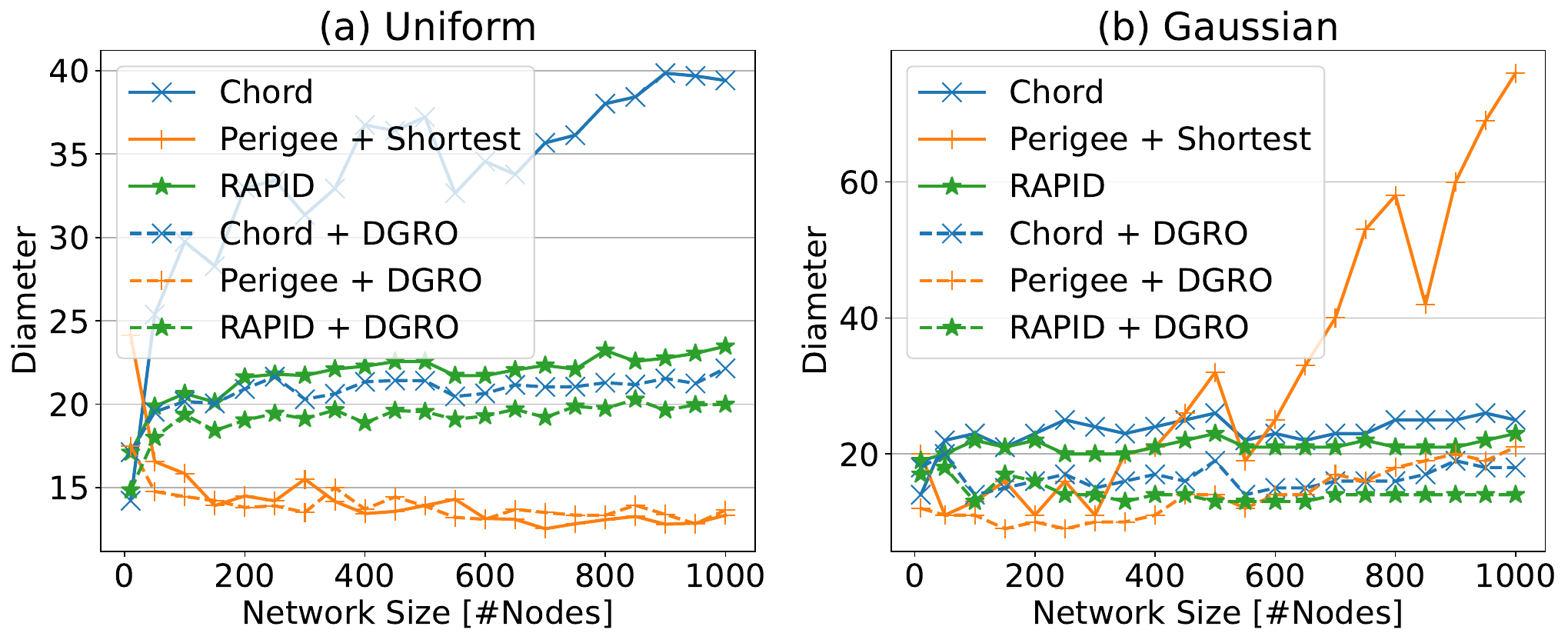}
    \caption{DGRO (dashed) reduces the diameter of CHORD (blue), Perigee (orange), and RAPID (green). Perigee is combined with a ring otherwise no connectivity guarantee.}
\label{fig:benchmark_synthetic_hamilton}
\vspace{-3mm}
\end{figure}

The choice between different rings is primarily dependent on the network's overall diameter when focusing specifically on this metric. Employing a single heuristic, such as minimizing path length through the shortest ring or random ring topology, may not always result in the most efficient outcomes in terms of diameter. Figure \ref{fig:benchmark_synthetic_hamilton} illustrates the DGRO helps reduce the network diameters of CHORD, PERIGEE, and RAPID, under two different latency distributions: uniform and Gaussian. In both uniform and Gaussian latency distributions, DGRO helps RAPID and Chord switch to a shortest ring from random topology, resulting in a better diameter. These configurations show a reduction in network diameter by more than 100\%. 

On the contrary, the DGRO helps Perigee adopt the random ring topology, mainly because the $\rho$ for Perigee is already very close to zero. The low value of $\rho_{Perigee}$ results in significantly large diameters, especially as the number of nodes approaches $1000$, proving to be highly unscalable. In such scenarios, DGRO helps the Perigee switches to a random ring and obtains a substantially low diameter of $20-30$. 

Hence, the single-heuristic approach can overlook other crucial aspects that contribute to an optimal network diameter, such as node distribution and link latencies, which a random ring might address by reducing the clustering of connections. Therefore, while the shortest ring topology may initially seem advantageous for minimizing diameter, it can lead to inefficiencies without considering the broader network context and the specific characteristics of the network's topology. We next present an ablation study on how different combinations of shortest ring and random ring impact the diameter.  
\begin{figure}[htp]
\vspace{-4mm}
    \centering
    \includegraphics[width=\linewidth]{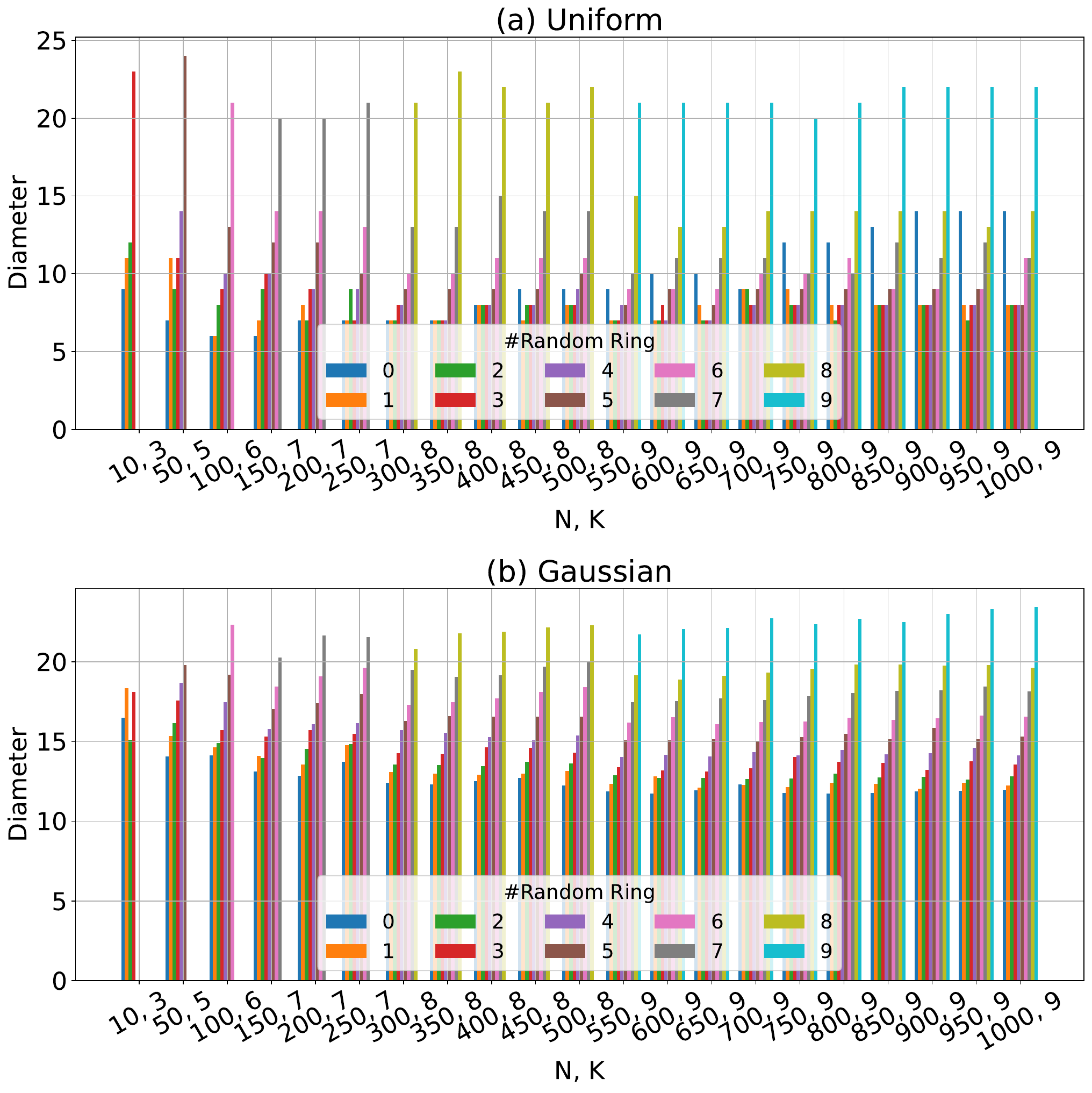}
    \caption{Benchmark how the number of DGRO rings impacts the diameter. Each color represents the number of random rings.}
    \label{fig:benchmark_synthetic_ablation}
    \vspace{-3mm}
\end{figure}
\subsubsection{Ablation study on the number of random rings}
RAPID is structured around $K$ random rings derived from $K$ consistent hashing functions. In Figure \ref{fig:benchmark_synthetic_ablation}, we modified up to $M$ of these rings to the shortest heuristic, with $M$ varying from 0 to $K$. The results show that for a uniform latency distribution, increasing $M$ does not significantly improve the diameter; notably, when the network size approaches 1000, the diameter dramatically increases when all rings are shortest rings. Conversely, for a Gaussian distribution, the diameter decreases monotonically as $M$ increases. Consequently, we developed DGRO to generalize across different latency distributions and network sizes, effectively adapting to the varying efficiencies observed under different configurations.

\subsubsection{DGRO}
\begin{figure}[htp]
\vspace{-4mm}
    \centering    \includegraphics[width=\linewidth]{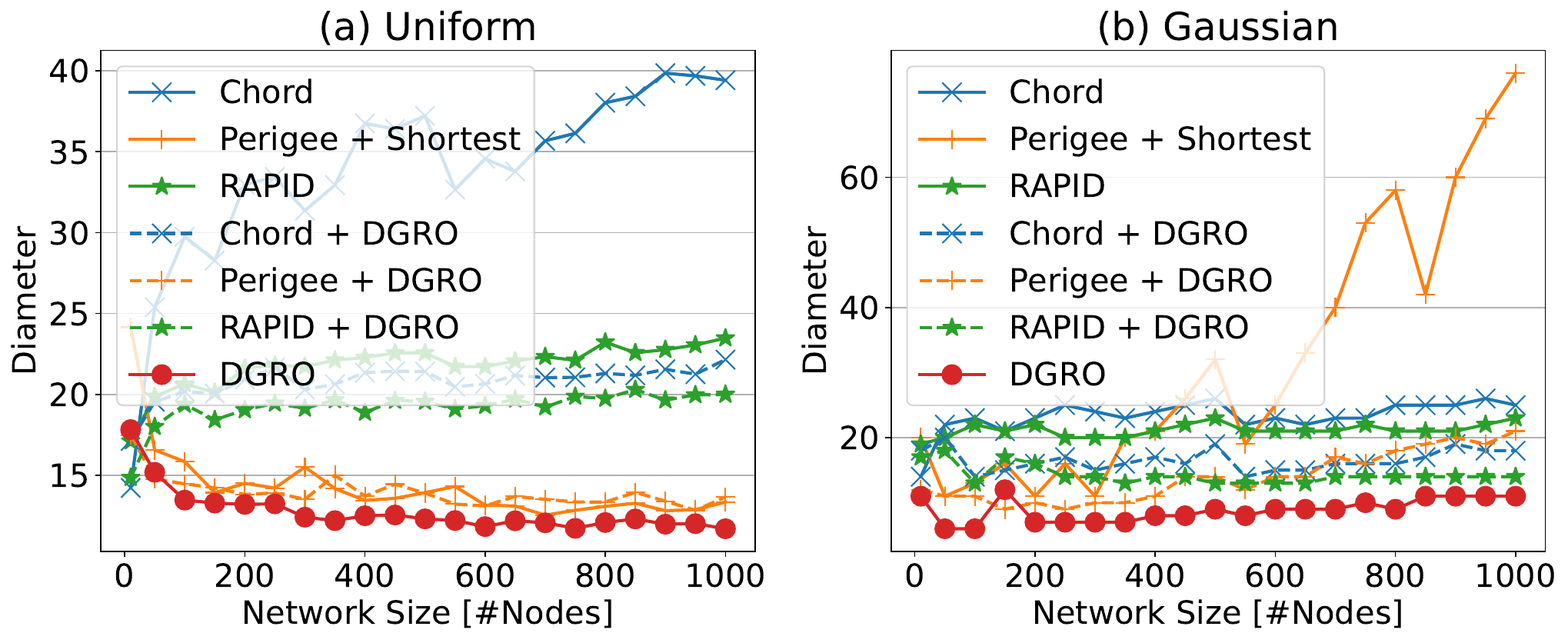}
    \caption{K-ring built by DGRO outperforms baselines.}
    \label{fig:benchmark_synthetic_all}
    \vspace{-4mm}
\end{figure}

Thanks to the capability of deep Q-learning to trade training time for solution quality, our DGRO model achieves a balance between random and shortest greedy approaches. As illustrated in Figure \ref{fig:benchmark_synthetic_all}, the DGRO topology (red solid line) outperforms all six baselines in terms of diameter for both uniform and Gaussian distribution. Furthermore, DGRO demonstrates excellent scalability; as the network size approaches 1000, the diameter remains consistently stable, indicating the model's robustness across larger network scales.

\subsubsection{Parallel-DGRO}
\begin{figure}
     \vspace{-3mm}
    \centering    \includegraphics[width=\linewidth]{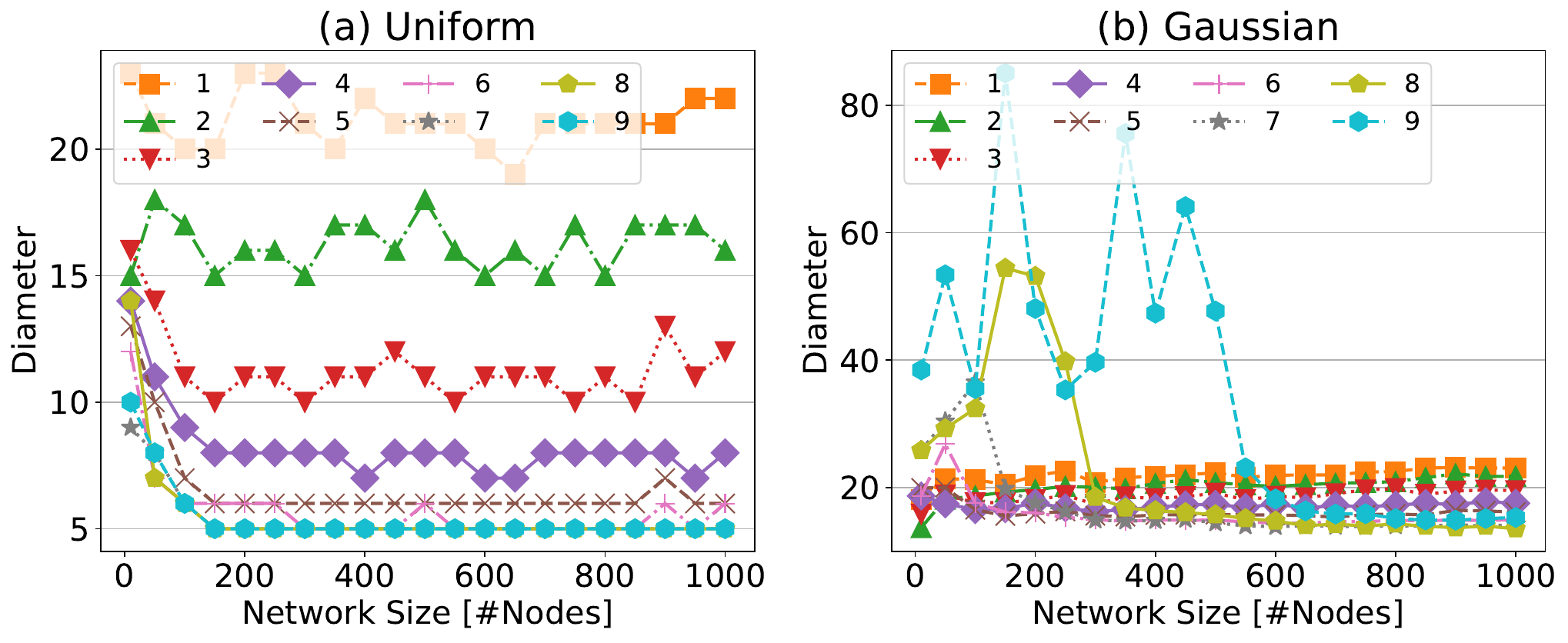}
    \caption{Parallel DGRO maintains the low diameter.}
\label{fig:benchmark_synthetic_distribute}
     \vspace{-3mm}
\end{figure}

In Figure \ref{fig:benchmark_synthetic_distribute}, we evaluate the performance of DGRO when constructed in parallel. In this setup, a random ring is initially segmented into $M$ partitions using a same stride, with each partition's starting node determined by a consistent hash function. In Figure \ref{fig:benchmark_synthetic_distribute}, the stride is set to $2^1 \sim 2^9$, Subsequently, these partitions concurrently reorder their internals using DGRO, while the construction within each partition is carried out sequentially. For both uniform and Gaussian distribution, the results show that even $8$-partition DGRO is comparable to that of the original DGRO, illustrating the efficiency of parallel DGRO construction in optimizing network topology. Next, we present the result on the realistic dataset, FABRIC, and Bitnode.

\subsection{Realistic Latency}
\label{sec:realisitic}
\subsubsection{Single heuristic ring}
\begin{figure}
    \centering
    \includegraphics[width=\linewidth]{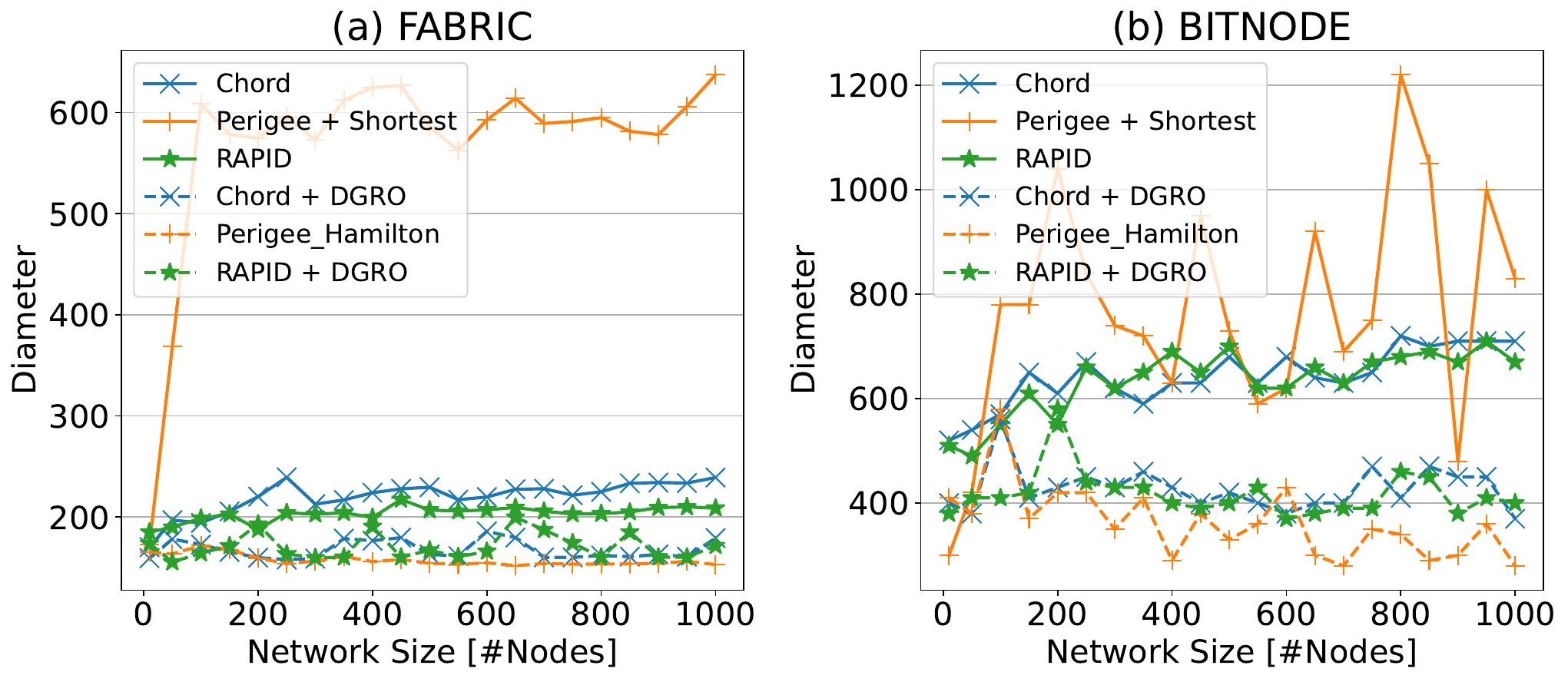}
    \caption{DGRO (dashed) reduces the diameter of CHORD (blue), Perigee (orange), and RAPID (green). Perigee is combined with a ring otherwise no connectivity guarantee.}
\label{fig:benchmark_real_hamilton}
     \vspace{-5mm}
\end{figure}

Figure \ref{fig:benchmark_real_hamilton} demonstrates how the ring selection can help reduce the diameter in two realistic datasets. For CHORD and RAPID, DGRO helps choose the shortest ring topology, which has lower diameter compared to a random ring. Conversely, in the case of Perigee, the random ring significantly outperforms the shortest ring.

\begin{figure}[t]
 \vspace{-3mm}
    \centering
    \includegraphics[width=\linewidth]{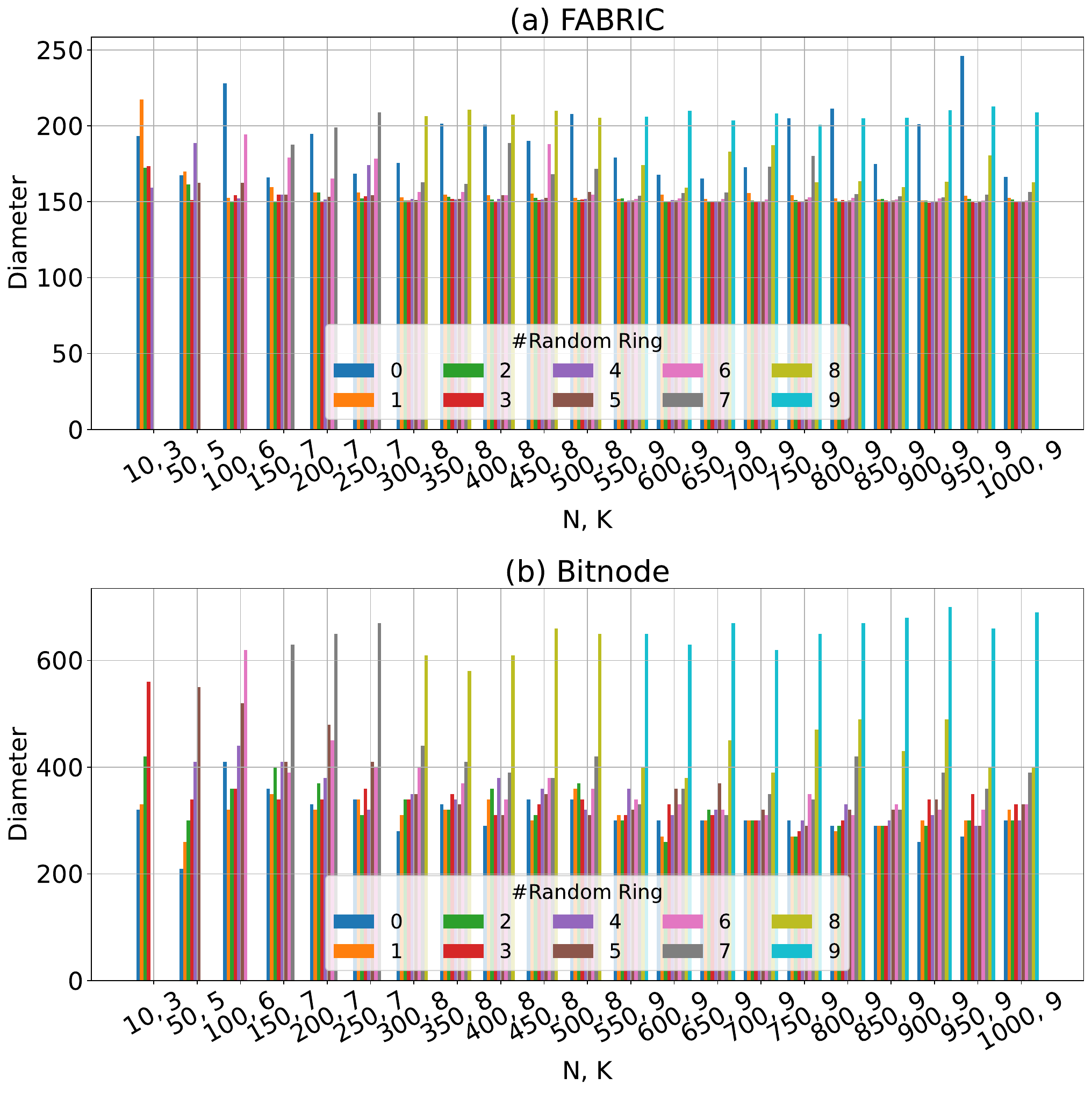}
    \caption{Benchmark how the number of DGRO rings impacts the diameter. Each color represents the number of random rings.}
\label{fig:benchmark_real_ablation}
        \vspace{-3mm}
\end{figure}
\subsubsection{Ablation study on the number of random rings}
As shown in Figure \ref{fig:benchmark_real_ablation}, within the RAPID framework that employs a hybrid heuristic combining \(M\) random rings with \((K-M)\) shortest rings, there is no consistent value of \(M\) that consistently yields the best network diameter across different latency distributions and network sizes. This observation indicates that the optimal configuration of \(M\) varies depending on the specific characteristics of the network and the latency distribution, suggesting a need for a more adaptive approach to configuring these hybrid topologies.

\subsubsection{DGRO}
\begin{figure}[htp]
    \centering
    \includegraphics[width=\linewidth]{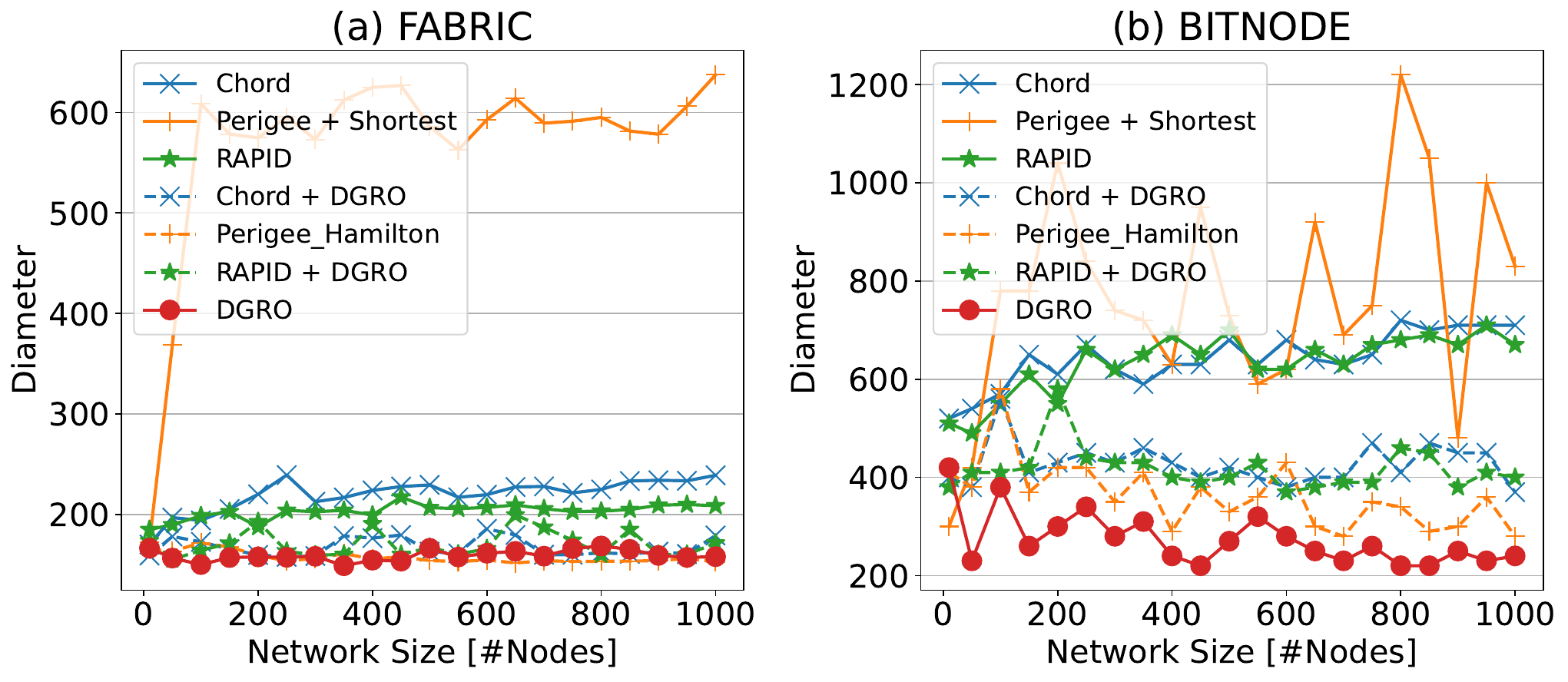}
    \caption{K-ring built by DGRO outperforms baselines.}
    \label{fig:benchmark_real_all}
    \vspace{0mm}
\end{figure}
Figure \ref{fig:benchmark_real_all} illustrated that DGRO consistently achieves a lower network diameter across a broad range of network sizes for both the FABRIC and Bitnode datasets compared to other methods. This performance demonstrates DGRO's effectiveness in optimizing network connectivity and efficiency in diverse settings.

\begin{figure}[h!]
    \vspace{-4mm}\centering\includegraphics[width=\linewidth]{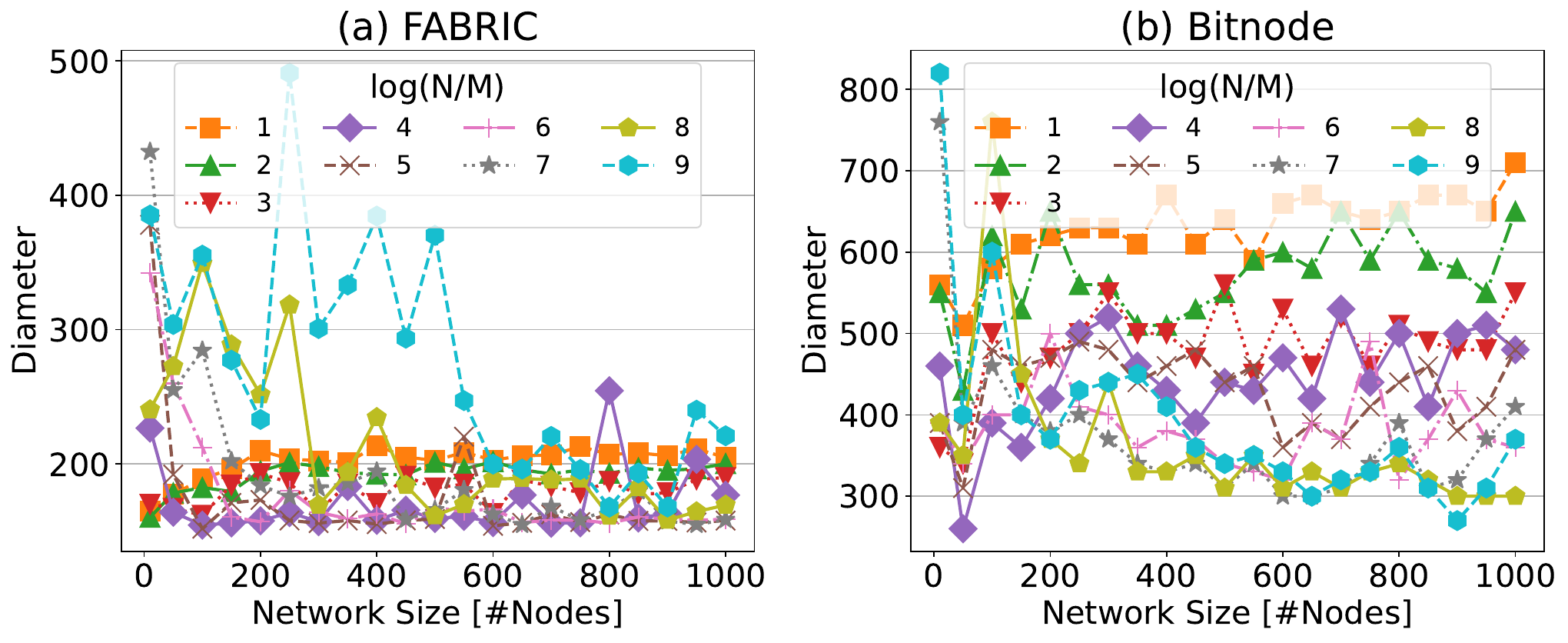}
    \caption{Parallel DGRO maintains the low diameter.}
    \label{fig:benchmark_real_distribute}
    \vspace{-4mm}
\end{figure}
\subsubsection{Parallel DGRO}
Figure \ref{fig:benchmark_real_distribute} tests the effectiveness of distributed construction on the FABRIC and Bitnode datasets. On the FABRIC dataset, using 2 to 32 partitions yields results similar to the original DGRO, demonstrating the model's robustness across different partitioning schemes. On the Bitnode dataset, maintaining four partitions achieves a diameter comparable to a single-partition setup, indicating effective scalability and network management with minimal partitioning.

\section{Conclusion}\label{sec:conclusion}
In this paper, we introduce \textbf{D}iameter-\textbf{G}uided \textbf{R}ing \textbf{O}ptimization (DGRO), which focuses on constructing rings with the smallest possible diameter, selecting the most effective ring configurations, and implementing these configurations in parallel. We first explore an integration of deep Q-learning and graph embedding to optimize the ring topology. We next propose a ring selection strategy that assesses the current topology's average latency against a global benchmark, facilitating integration into modern peer-to-peer protocols and substantially reducing network diameter. To further enhance scalability, we propose a parallel strategy that distributes the topology construction process into separate partitions simultaneously. In the future, we will integrate DGRO into a real system and evaluate its performance in a real-world environment, such as Slurm \cite{yoo2003slurm}. Besides, we will explore the online update of DGRO.

\newpage
\renewcommand*{\bibfont}{\footnotesize}
\printbibliography[]

@article{wu2024turbofft,
  title={TurboFFT: A High-Performance Fast Fourier Transform with Fault Tolerance on GPU},
  author={Wu, Shixun and Zhai, Yujia and Liu, Jinyang and Huang, Jiajun and Jian, Zizhe and Dai, Huangliang and Di, Sheng and Chen, Zizhong and Cappello, Franck},
  journal={arXiv preprint arXiv:2405.02520},
  year={2024}
}

@article{liu2024high,
  title={High-performance Effective Scientific Error-bounded Lossy Compression with Auto-tuned Multi-component Interpolation},
  author={Liu, Jinyang and Di, Sheng and Zhao, Kai and Liang, Xin and Jin, Sian and Jian, Zizhe and Huang, Jiajun and Wu, Shixun and Chen, Zizhong and Cappello, Franck},
  journal={Proceedings of the ACM on Management of Data},
  volume={2},
  number={1},
  pages={1--27},
  year={2024},
  publisher={ACM New York, NY, USA}
}

@inproceedings{liu2023stationary,
  title={Stationary deep reinforcement learning with quantum k-spin hamiltonian regularization},
  author={Liu, Xiao-Yang and Li, Zechu and Wu, Shixun and Wang, Xiaodong},
  booktitle={ICLR 2023 Workshop on Physics for Machine Learning},
  year={2023}
}

@inproceedings{wu2023ft,
  title={Ft-gemm: A fault tolerant high performance gemm implementation on x86 cpus},
  author={Wu, Shixun and Zhai, Yujia and Huang, Jiajun and Jian, Zizhe and Chen, Zizhong},
  booktitle={Proceedings of the 32nd International Symposium on High-Performance Parallel and Distributed Computing},
  pages={323--324},
  year={2023}
}

@inproceedings{jian2024cliz,
  title={CliZ: Optimizing lossy compression for climate datasets with adaptive fine-tuned data prediction},
  author={Jian, Zizhe and Di, Sheng and Liu, Jinyang and Zhao, Kai and Liang, Xin and Xu, Haiying and Underwood, Robert and Wu, Shixun and Huang, Jiajun and Chen, Zizhong and others},
  booktitle={2024 IEEE International Parallel and Distributed Processing Symposium (IPDPS)},
  pages={417--429},
  year={2024},
  organization={IEEE}
}

@inproceedings{huang2023exploring,
  title={Exploring Wavelet Transform Usages for Error-bounded Scientific Data Compression},
  author={Huang, Jiajun and Liu, Jinyang and Di, Sheng and Zhai, Yujia and Jian, Zizhe and Wu, Shixun and Zhao, Kai and Chen, Zizhong and Guo, Yanfei and Cappello, Franck},
  booktitle={2023 IEEE International Conference on Big Data (BigData)},
  pages={4233--4239},
  year={2023},
  organization={IEEE}
}

@article{johnston2023curriculum,
  title={A Curriculum Learning Approach to Optimization with Application to Downlink Beamforming},
  author={Johnston, Jeremy and Liu, Xiao-Yang and Wu, Shixun and Wang, Xiaodong},
  journal={IEEE Transactions on Signal Processing},
  year={2023},
  publisher={IEEE}
}

@article{liu2023cusz,
  title={cuSZ-I: High-Fidelity Error-Bounded Lossy Compression for Scientific Data on GPUs},
  author={Liu, Jinyang and Tian, Jiannan and Wu, Shixun and Di, Sheng and Zhang, Boyuan and Huang, Yafan and Zhao, Kai and Li, Guanpeng and Tao, Dingwen and Chen, Zizhong and others},
  journal={arXiv preprint arXiv:2312.05492},
  year={2023}
}

@inproceedings{johnston2023downlink,
  title={Downlink Beamforming Optimization via Deep Learning},
  author={Johnston, Jeremy and Liu, Xiao-Yang and Wu, Shixun and Wang, Xiaodong},
  booktitle={2023 59th Annual Allerton Conference on Communication, Control, and Computing (Allerton)},
  pages={1--5},
  year={2023},
  organization={IEEE}
}

@inproceedings{wu2023anatomy,
  title     = {Anatomy of High-Performance GEMM with Online Fault Tolerance on GPUs},
  author    = {Wu, Shixun and Zhai, Yujia and Liu, Jinyang and Huang, Jiajun and Jian, Zizhe and Wong, Bryan and Chen, Zizhong},
  booktitle = {Proceedings of the 37th International Conference on Supercomputing},
  pages     = {360--372},
  year      = {2023}
}

@inproceedings{adriaens2022diameter,
  title={Diameter minimization by shortcutting with degree constraints},
  author={Adriaens, Florian and Gionis, Aristides},
  booktitle={2022 IEEE International Conference on Data Mining (ICDM)},
  pages={843--848},
  year={2022},
  organization={IEEE}
}

@article{stoica2001chord,
  title={Chord: A scalable peer-to-peer lookup service for internet applications},
  author={Stoica, Ion and Morris, Robert and Karger, David and Kaashoek, M Frans and Balakrishnan, Hari},
  journal={ACM SIGCOMM computer communication review},
  volume={31},
  number={4},
  pages={149--160},
  year={2001},
  publisher={ACM New York, NY, USA}
}

@inproceedings{mao2020perigee,
  title={Perigee: {Efficient} {Peer}-to-{Peer} {Network} {Design} for {Blockchains}},
  author={Mao, Yifan and Deb, Soubhik and Venkatakrishnan, Shaileshh Bojja and Kannan, Sreeram and Srinivasan, Kannan},
  booktitle={Proceedings of the 39th Symposium on Principles of Distributed Computing},
  pages={428--437},
  year={2020}
}

@article{khalil2017learning,
  title={{Learning} {Combinatorial} {Optimization} {Algorithms} over {Graphs}},
  author={Dai, Hanjun and Khalil, Elias and Zhang, Yuyu and Dilkina, Bistra and Song, Le},
  journal={Advances in neural information processing systems},
  volume={30},
  year={2017}
}

@techreport{none2021toward,
  title={Toward a Seamless Integration of Computing, Experimental, and Observational Science Facilities: A Blueprint to Accelerate Discovery},
  author={None, None},
  year={2021},
  institution={USDOE Office of Science (SC), Washington, DC (United States). Advanced~…}
}

@inproceedings{brown2021vision,
  title={A vision for the ascr facilities enterprise},
  author={Brown, Benjamin},
  booktitle={Meeting of the Advanced Scientific Computing Advisory Committee},
  year={2021}
}

@techreport{ahrens2022envisioning,
  title={Envisioning Science in 2050},
  author={Ahrens, James and Boehnlein, Amber and Carlson, Rich and Elliot, Joshua and Fagnan, Kjiersten and Ferrier, Nicola and Foster, Ian and Gimpel, Lee and Shalf, John and Ratner, Dan},
  year={2022},
  institution={USDOE Office of Science (SC)(United States)}
}

@article{bard2022lbnl,
  title={The LBNL superfacility project report},
  author={Bard, Deborah and Snavely, Cory and Gerhardt, Lisa and Lee, Jason and Totzke, Becci and Antypas, Katie and Arndt, William and Blaschke, Johannes and Byna, Suren and Cheema, Ravi and others},
  journal={arXiv preprint arXiv:2206.11992},
  year={2022}
}

@inproceedings{sridharan2012study,
  title={A study of DRAM failures in the field},
  author={Sridharan, Vilas and Liberty, Dean},
  booktitle={SC'12: Proceedings of the International Conference on High Performance Computing, Networking, Storage and Analysis},
  pages={1--11},
  year={2012},
  organization={IEEE}
}

@inproceedings{patterson1988case,
  title={A case for redundant arrays of inexpensive disks (RAID)},
  author={Patterson, David A and Gibson, Garth and Katz, Randy H},
  booktitle={Proceedings of the 1988 ACM SIGMOD international conference on Management of data},
  pages={109--116},
  year={1988}
}

@inproceedings{park2009reliability,
  title={Reliability and performance enhancement technique for SSD array storage system using RAID mechanism},
  author={Park, Kwanghee and Lee, Dong-Hwan and Woo, Youngjoo and Lee, Geunhyung and Lee, Ju-Hong and Kim, Deok-Hwan},
  booktitle={2009 9th International Symposium on Communications and Information Technology},
  pages={140--145},
  year={2009},
  organization={IEEE}
}

@article{losada2020fault,
  title={Fault tolerance of MPI applications in exascale systems: The ULFM solution},
  author={Losada, Nuria and Gonz{\'a}lez, Patricia and Mart{\'\i}n, Mar{\'\i}a J and Bosilca, George and Bouteiller, Aur{\'e}lien and Teranishi, Keita},
  journal={Future Generation Computer Systems},
  volume={106},
  pages={467--481},
  year={2020},
  publisher={Elsevier}
}

@article{stelling1999fault,
  title={A fault detection service for wide area distributed computations},
  author={Stelling, Paul and DeMatteis, Cheryl and Foster, Ian and Kesselman, Carl and Lee, Craig and von Laszewski, Gregor},
  journal={Cluster Computing},
  volume={2},
  pages={117--128},
  year={1999},
  publisher={Springer}
}

@book{barroso2022datacenter,
  title={The datacenter as a computer: An introduction to the design of warehouse-scale machines},
  author={Barroso, Luis Andre and Clidaras, Jimmy},
  year={2022},
  publisher={Springer Nature}
}

@article{dean2013tail,
  title={The tail at scale},
  author={Dean, Jeffrey and Barroso, Luiz Andr{\'e}},
  journal={Communications of the ACM},
  volume={56},
  number={2},
  pages={74--80},
  year={2013},
  publisher={ACM New York, NY, USA}
}

@article{cappello2014toward,
  title={Toward exascale resilience: 2014 update},
  author={Cappello, Franck and Al, Geist and Gropp, William and Kale, Sanjay and Kramer, Bill and Snir, Marc},
  journal={Supercomputing Frontiers and Innovations: an International Journal},
  volume={1},
  number={1},
  pages={5--28},
  year={2014},
  publisher={South Ural State University Chelyabinsk, Russia, Russia}
}

@misc{zookeeper2010apache,
  title={Apache zookeeper},
  author={ZooKeeper, Apache},
  year={2010}
}

@misc{etcd2014,
  author = "{ETCD}",
  title = "{etcd}",
  year = {2014},
  howpublished = {\url{https://github.com/coreos/etcd}},
}

@misc{netflix2014,
  author = "{NETFLIX}",
  title = "{netflix}",
  year = {2014},
  howpublished = {\url{https://github.com/Netflix/eureka}},
}

@inproceedings{yoo2003slurm,
  title={Slurm: Simple linux utility for resource management},
  author={Yoo, Andy B and Jette, Morris A and Grondona, Mark},
  booktitle={Workshop on job scheduling strategies for parallel processing},
  pages={44--60},
  year={2003},
  organization={Springer}
}

@inproceedings{burrows2006chubby,
  title={The Chubby lock service for loosely-coupled distributed systems},
  author={Burrows, Mike},
  booktitle={Proceedings of the 7th symposium on Operating systems design and implementation},
  pages={335--350},
  year={2006}
}

@article{kelley2014eureka,
  title={Eureka! why you shouldn’t use zookeeper for service discovery},
  author={Kelley, Peter},
  journal={Dosegljivo: https://tech. knewton. com/blog/2014/12/eureka-shouldnt-use-zookeeper-service-discovery},
  year={2014}
}

@book{hewitt2010cassandra,
  title={Cassandra: the definitive guide},
  author={Hewitt, Eben},
  year={2010},
  publisher={" O'Reilly Media, Inc."}
}

@misc{akka2009,
  author = "{TYPESAFE}",
  title = "{Akka}",
  year = {2009},
  howpublished = {\url{http://akka.io/}},
}

@misc{redis2009,
  author = "{REDIS}",
  title = "{Redis}",
  year = {2009},
  howpublished = {\url{http://redis.io/}},
}

@inproceedings{newell2016optimizing,
  title={Optimizing distributed actor systems for dynamic interactive services},
  author={Newell, Andrew and Kliot, Gabriel and Menache, Ishai and Gopalan, Aditya and Akiyama, Soramichi and Silberstein, Mark},
  booktitle={Proceedings of the Eleventh European Conference on Computer Systems},
  pages={1--15},
  year={2016}
}

@inproceedings{suresh2018stable,
  title={Stable and consistent membership at scale with rapid},
  author={Suresh, Lalith and Malkhi, Dahlia and Gopalan, Parikshit and Carreiro, Ivan Porto and Lokhandwala, Zeeshan},
  booktitle={2018 USENIX Annual Technical Conference (USENIX ATC 18)},
  pages={387--400},
  year={2018}
}

@misc{scylladb2013,
  author = "{SCYLLA}",
  title = "{ScyllaDB}",
  year = {2013},
  howpublished = {\url{http://www.scylladb.com/}},
}

@inproceedings{van1998gossip,
  title={A gossip-style failure detection service},
  author={Van Renesse, Robbert and Minsky, Yaron and Hayden, Mark},
  booktitle={Middleware’98: IFIP International Conference on Distributed Systems Platforms and Open Distributed Processing},
  pages={55--70},
  year={1998},
  organization={Springer}
}

@article{van2003astrolabe,
  title={Astrolabe: A robust and scalable technology for distributed system monitoring, management, and data mining},
  author={Van Renesse, Robbert and Birman, Kenneth P and Vogels, Werner},
  journal={ACM transactions on computer systems (TOCS)},
  volume={21},
  number={2},
  pages={164--206},
  year={2003},
  publisher={ACM New York, NY, USA}
}

@inproceedings{das2002swim,
  title={Swim: Scalable weakly-consistent infection-style process group membership protocol},
  author={Das, Abhinandan and Gupta, Indranil and Motivala, Ashish},
  booktitle={Proceedings International Conference on Dependable Systems and Networks},
  pages={303--312},
  year={2002},
  organization={IEEE}
}

@misc{mccaffrey2015,
  author = {McCaffrey, C.},
  title = {Building Scalable Stateful Services},
  year = {2015},
  howpublished = {\url{https://speakerdeck.com/caitiem20/building-scalable-stateful-services}},
  note = {Accessed: 2024-10-10}
}

@article{benini2002networks,
  title={Networks on chips: A new SoC paradigm},
  author={Benini, Luca and De Micheli, Giovanni},
  journal={computer},
  volume={35},
  number={1},
  pages={70--78},
  year={2002},
  publisher={IEEE}
}

@inproceedings{laoutaris2008bounded,
  title={Bounded budget connection (BBC) games or how to make friends and influence people, on a budget},
  author={Laoutaris, Nikolaos and Poplawski, Laura J and Rajaraman, Rajmohan and Sundaram, Ravi and Teng, Shang-Hua},
  booktitle={Proceedings of the twenty-seventh ACM symposium on Principles of distributed computing},
  pages={165--174},
  year={2008}
}

@inproceedings{demaine2010minimizing,
  title={Minimizing the diameter of a network using shortcut edges},
  author={Demaine, Erik D and Zadimoghaddam, Morteza},
  booktitle={Scandinavian Workshop on Algorithm Theory},
  pages={420--431},
  year={2010},
  organization={Springer}
}

@inproceedings{meyerson2009minimizing,
  title={Minimizing average shortest path distances via shortcut edge addition},
  author={Meyerson, Adam and Tagiku, Brian},
  booktitle={International Workshop on Approximation Algorithms for Combinatorial Optimization},
  pages={272--285},
  year={2009},
  organization={Springer}
}

@article{tan2017shortcutting,
  title={Shortcutting directed and undirected networks with a degree constraint},
  author={Tan, Richard B and Van Leeuwen, Erik Jan and Van Leeuwen, Jan},
  journal={Discrete Applied Mathematics},
  volume={220},
  pages={91--117},
  year={2017},
  publisher={Elsevier}
}

@article{chepoi2002augmenting,
  title={Augmenting trees to meet biconnectivity and diameter constraints},
  author={Chepoi, Victor and Vaxes, Yann},
  journal={Algorithmica},
  volume={33},
  pages={243--262},
  year={2002},
  publisher={Springer}
}

@article{li1992minimum,
  title={On the minimum-cardinality-bounded-diameter and the bounded-cardinality-minimum-diameter edge addition problems},
  author={Li, Chung-Lun and McCormick, S Thomas and Simchi-Levi, David},
  journal={Operations Research Letters},
  volume={11},
  number={5},
  pages={303--308},
  year={1992},
  publisher={Elsevier}
}

@article{chung1984diameter,
  title={Diameter bounds for altered graphs},
  author={Chung, Fan RK and Garey, Michael R},
  journal={Journal of graph theory},
  volume={8},
  number={4},
  pages={511--534},
  year={1984},
  publisher={Wiley Online Library}
}

@article{bokhari1986reducing,
  title={Reducing the diameters of computer networks},
  author={Bokhari and Raza},
  journal={IEEE transactions on computers},
  volume={100},
  number={8},
  pages={757--761},
  year={1986},
  publisher={IEEE}
}

@inproceedings{garimella2017reducing,
  title={Reducing controversy by connecting opposing views},
  author={Garimella, Kiran and De Francisci Morales, Gianmarco and Gionis, Aristides and Mathioudakis, Michael},
  booktitle={Proceedings of the tenth ACM international conference on web search and data mining},
  pages={81--90},
  year={2017}
}

@inproceedings{haddadan2021repbublik,
  title={Repbublik: Reducing polarized bubble radius with link insertions},
  author={Haddadan, Shahrzad and Menghini, Cristina and Riondato, Matteo and Upfal, Eli},
  booktitle={Proceedings of the 14th ACM International Conference on Web Search and Data Mining},
  pages={139--147},
  year={2021}
}

@article{interian2021polarization,
  title={Polarization reduction by minimum-cardinality edge additions: Complexity and integer programming approaches},
  author={Interian, Ruben and Moreno, Jorge R and Ribeiro, Celso C},
  journal={International Transactions in Operational Research},
  volume={28},
  number={3},
  pages={1242--1264},
  year={2021},
  publisher={Wiley Online Library}
}

@article{vinyals2015pointer,
  title={Pointer networks},
  author={Vinyals, Oriol and Fortunato, Meire and Jaitly, Navdeep},
  journal={Advances in neural information processing systems},
  volume={28},
  year={2015}
}

@article{bello2016neural,
  title={Neural combinatorial optimization with reinforcement learning},
  author={Bello, Irwan and Pham, Hieu and Le, Quoc V and Norouzi, Mohammad and Bengio, Samy},
  journal={arXiv preprint arXiv:1611.09940},
  year={2016}
}

@article{graves2016hybrid,
  title={Hybrid computing using a neural network with dynamic external memory},
  author={Graves, Alex and Wayne, Greg and Reynolds, Malcolm and Harley, Tim and Danihelka, Ivo and Grabska-Barwi{\'n}ska, Agnieszka and Colmenarejo, Sergio G{\'o}mez and Grefenstette, Edward and Ramalho, Tiago and Agapiou, John and others},
  journal={Nature},
  volume={538},
  number={7626},
  pages={471--476},
  year={2016},
  publisher={Nature Publishing Group UK London}
}

@article{wu2024ft,
  title={FT K-Means: A High-Performance K-Means on GPU with Fault Tolerance},
  author={Wu, Shixun and Ding, Yitong and Zhai, Yujia and Liu, Jinyang and Huang, Jiajun and Jian, Zizhe and Dai, Huangliang and Di, Sheng and Wong, Bryan M and Chen, Zizhong and others},
  journal={arXiv preprint arXiv:2408.01391},
  year={2024}
}

%\newpage
% \setcounter{tocdepth}{4}
%\renewcommand{\contentsname}{ToC, not part of the submission}
%{\footnotesize\sffamily\tableofcontents}

\end{document}